\documentclass[12pt,preprint]{aastex}
\usepackage{graphics}

\def\gtorder{\mathrel{\raise.3ex\hbox{$>$}\mkern-14mu
             \lower0.6ex\hbox{$\sim$}}}
\def\ltorder{\mathrel{\raise.3ex\hbox{$<$}\mkern-14mu
             \lower0.6ex\hbox{$\sim$}}}

\def\Msun{\>{\rm M_{\odot}}}

\shorttitle{Scattered Light Spectrum of TW Hya}
\shortauthors{Debes et al.}

\begin{document}
\title{The 0.5-2.22\micron\ Scattered Light Spectrum of the Disk Around TW Hya: Detection of a Partially Filled Disk Gap at 80~AU\footnote{Based on observations made with the NASA/ESA Hubble Space Telescope, obtained at the Space Telescope Science Institute, which is operated by the Association of Universities for Research in Astronomy, Inc., under NASA contract NAS 5-26555. These observations are associated with program \#'s 10167, 8624, 7226, and 7233.}}
\author{John H.Debes\altaffilmark{1}, Hannah Jang-Condell\altaffilmark{2},Alycia J. Weinberger\altaffilmark{3}, Aki Roberge\altaffilmark{4}, Glenn Schneider\altaffilmark{5}}

\altaffiltext{1}{Space Telescope Science Institute, Baltimore, MD 21218}
\altaffiltext{2}{University of Wyoming, Laramie, WY 82071}
\altaffiltext{3}{Department of Terrestrial Magnetism, Carnegie Institution of Washington, Washington, D.C. 20015}
\altaffiltext{4}{Goddard Space Flight Center, Greenbelt, MD 20771}
\altaffiltext{5}{Steward Observatory, The University of Arizona, Tuscon, AZ 85721}

\begin{abstract}
We present a 0.5-2.2~\micron\ scattered light spectrum of the circumstellar disk
around TW Hya from a combination of spatially resolved HST STIS spectroscopy 
and NICMOS
coronagraphic images of the disk.  We investigate the morphology of the disk at distances $>$ 40~AU over this wide range of wavelengths, and identify the presence of a depression in surface brightness at $\sim$80~AU that could be caused by a gap in the disk.   Additionally, we quantify the surface brightness, azimuthal symmetry, and spectral character of the disk as a function of radius.  Our analysis shows that the scattering efficiency of the dust is largely neutral to blue over the observed wavelengths.   We model the disk as a steady $\alpha$-disk with an ad hoc gap structure.  The thermal properties of the disk are self-consistently calculated using a three-dimensional radiative transfer code that uses ray-tracing to model the heating of the disk interior and scattered light images.  We find a good fit to the data over a wide range of distances from the star if we use a model disk with a partially filled gap of 30$\%$ depth at 80~AU and with a self-similar truncation knee at 100~AU.  The origin of the gap is unclear, but it could arise from a transition in the nature of the disk's dust composition or the presence of a planetary companion.  Based on scalings to previous hydrodynamic simulations of gap opening criteria for embedded proto-planets, we estimate that a planetary companion forming the gap could have a mass between 6-28~$M_\oplus$.  

\end{abstract}  


\section{Introduction}
TW Hya is the closest example of a star with a young, gas-rich protoplanetary
disk. As such, it is an ideal target to study how disk evolution is
coupled to planet formation because the disk can be observed at high spatial
resolution and with great sensitivity.   First detected from its infrared excess
\citep{Rucinski:1983}, the disk emits strongly at wavelengths longer than 3~$\mu$m 
with a peak in its spectral energy distribution between 10 and 100~$\mu$m \citep{Weinberger:2002}.

TW Hya may not be typical for its age. It is part of the eponymously named TW
Hydrae Association, a collection of about two dozen stars, amongst which it
sports the most massive, gas-rich disk of the group. TW Hya is 54 $\pm$ 6 pc away from Earth in
the new \citet{vanleeuwen:2007} Hipparcos catalog.  The age of the TW Hya
association is found from a variety of studies of the ensemble of stars,
including measurements of the dynamics (8.3 $\pm$ 0.3 Myr)
\citep{delaReza:2006}, lithium depletion boundary (12 $\pm$ 8 Myr)
\citep{Mentuch:2008}, and these in combination with pre-main sequence tracks
(10$^{+10}_{-7}$ Myr) \citep{BarradoyNavascues:2006}. The TW Hya stars may not
be coeval, and TW Hya could even be at the young end of their age distribution
\citep{Weinberger:2013}.  Associations just a bit older than TW Hya, such as
$\beta$ Pic, have no optically thick, massive protoplanetary disks.

Despite reservations about its representativeness, intensive studies at nearly every
wavelength and spectral resolution have been used to try to understand the TW
Hya disk's structure and composition. We summarize some of the main findings
here.  Resolved emission at 7mm indicated the central disk is mostly clear inside of
4 AU \citep{Calvet:2002,Hughes:2007}.  However, 2 $\mu$m emission resolved by the Keck
Interferometer show that an optically thin dust disk comprised of small grains
comes to within 0.06 AU of the star \citep{akeson11}.  The spectral energy distribution,
particularly the large amount of emission at submm through cm wavelengths
\citep{Weintraub:1989,Wilner:2000,Weinberger:2002,Wilner:2005,andrews12} indicates substantial grain growth
to at least cm sizes.  Resolved CO line maps show the disk in Keplerian
rotation with an inclination of 7 $\pm$ 1$^\circ$ beyond 50 AU \citep{qi04,andrews12,hughes11}.

The disk has been spatially resolved in scattered light in the visible and
near-infrared \citep{Krist:2000,Weinberger:2002,apai04,Roberge:2005}.  These show that
the optically thick disk extends to at least 280 AU.  They also show
asymmetries and changes in surface brightness of the disk in the inner 150 AU.
 The presence of both small and large
grains throughout the 4 AU - 200 AU disk and the presence of small grains in a
region where they should quickly be removed, suggests that they are being
regenerated through collisions at multiple locations.

The disk chemistry has also been probed in a spatially resolved manner in
submm lines \citep{Qi:2006, Qi:2008}.  For example, there appears to be ongoing
deuterium fractionation in the outer disk, and that suggests pristine
nebular material is not preserved, as is often assumed for comets.

We have undertaken a multiwavelength visible to near-infrared study of
scattered light from the disk in order to address the structure and composition of 
the TW Hya disk. Observations in the visible to
near-infrared can detect ices and silicates due to their broad absorption
features and organics due to their red slopes.  These types of observations
are routinely applied to comets and Kuiper Belt objects, which are thought to
be the planetesimal remnants of circumstellar disks.  In addition, the
spectral scattering efficiency of disk dust grains can constrain their grain sizes at
the disk surface. The mixture of dust grains should reflect a combination of
vertical mixing from the midplane (where large grains are presumably formed), 
radial transport, and collisions.  Finally but by no means least, we wish to
understand the vertical structure of the disk and whether it shows evidence
for forming planets.

\section{Observations}
We took coronagraphic images with the F171M, F180M, F204M, and F222M filters (central $\lambda$=1.71, 1.80, 2.04, and 2.22~\micron\ respectively) on 09 May 2005 with the NICMOS
camera 2 for TW Hya and the PSF reference star, CD-43$^\circ$2742 as part of Program GO 10167.  The observations include direct
images of both stars outside the coronagraphic hole with short exposures for point source photometry, as 
well as longer exposures for coronagraphic high contrast imaging. 
The instrumentally calibrated and reduced images discussed in this paper
were created from the raw NICMOS {\it multiaccum} exposures following
the processing methodolgy  described by \S3 of \citet{schneider05} and 
references therein.

For photometric analysis, each calibrated direct image was used to determine the total photometry of the star and 
empirically determine the scaling ratios between TW~Hya and the PSF reference in each filter band.  The three images for
each star in each filter were located at different positions on the 
detector.
We used a median combination of the three dither points to create a final image
of each star to derive a ratio for scaling and for photometry of
TW Hya.  We used a 16.5 pixel radius circular aperture to determine the
photometry.  The background in the images is zero, so no background annulus was used.  The individual dither points were used to get a rough estimate of
the uncertainty in the ratios and photometry.  Table \ref{tab:scalings} lists the photometry of TW Hya in each band with uncertainties and the scaling ratios for each filter. 

In order to determine the best subtraction we minimized a chi-squared metric on a 
region of the target image dominated by the star's diffraction spikes.  We assumed that good
subtraction of the diffraction spikes corresponded to the best subtraction
of the PSF within the region of interest \citep{cnc03}.  We iteratively created subtractions for combinations of scaling and pixel offsets until we found an image that produced the lowest chi-squared measure.  We searched within 1-$\sigma$ of the scaling ratios as determined by Table \ref{tab:scalings} and within $\pm$1~pixel to find the best x and y pixel offsets. 

To quantify the systematic effects on the photometry, we repeated the subtractions varying the PSF scalings and offsets by $\pm$1 $\sigma$ from the minimum chi-square solution found above. Using a circular photometric aperture matched to the size of the disk and avoiding masked areas due to diffraction spikes, we found the standard deviation in the disk flux densities from this suite of subtractions.  We then propagated this uncertainty into the total uncertainty in the flux density of the disk per pixel.

\begin{figure}
\plotone{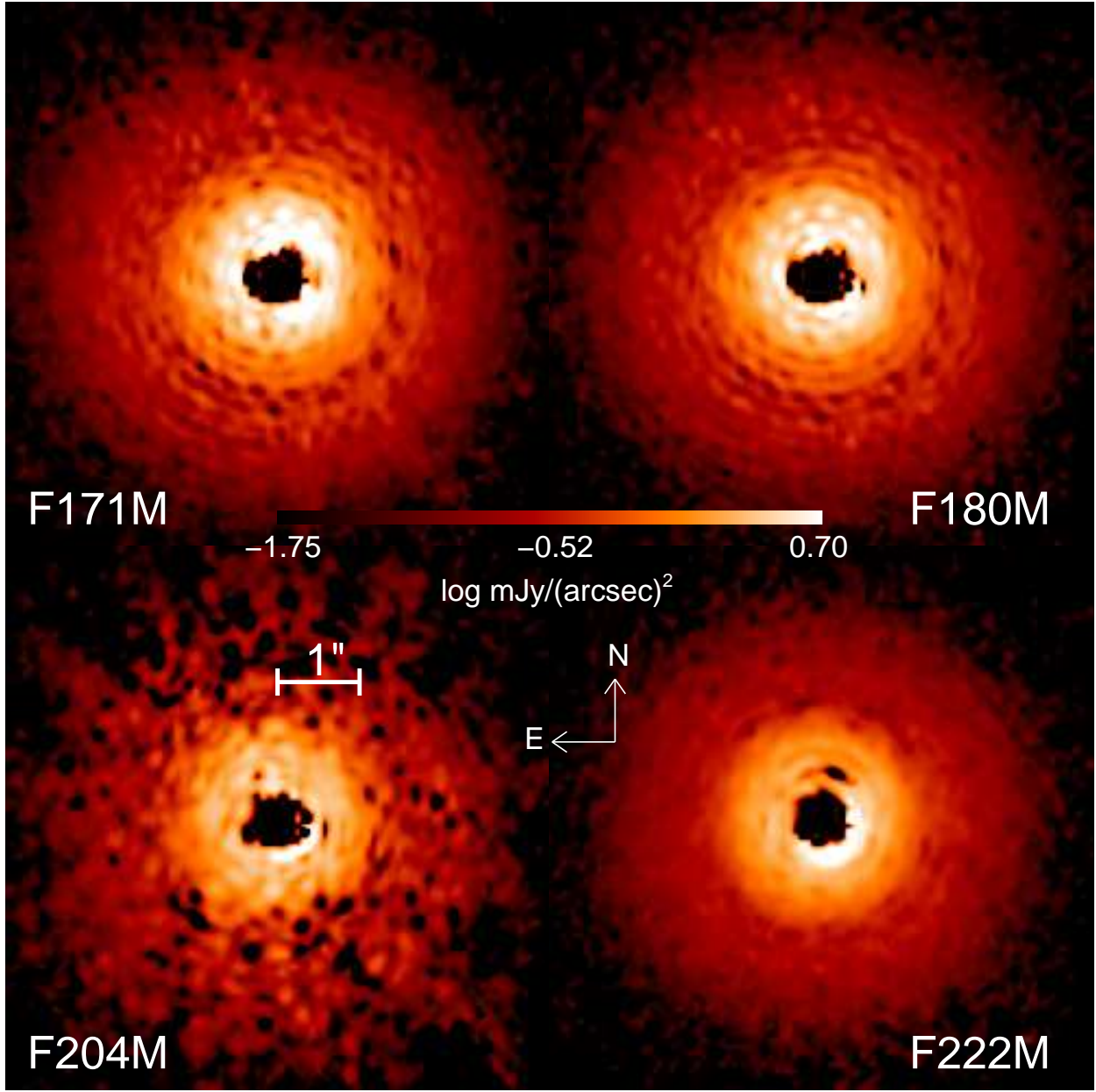}
\caption{\label{fig:f1} Final, roll combined, PSF-subtracted images of TW Hya in the four
medium band filters observed, F171M, F180M, F204M, and F222M.  Regions in which there is no data, i.e. under the coronagraph are blacked out.  The mottled appearance in the filters, particularly F204M, is the result of systematic residuals from diffraction and scattering patterns within the NICMOS camera.}
\end{figure}

We observed TW Hya in each medium band filter at two distinct spacecraft orientations or roll angles separated by 28$^\circ$.  This is essentially an azimuthal dither that allows true
objects within the field of view to be distinguished with instrumental artifacts
that do not rotate with a change in orientation.  The reference star, CD-43$^\circ$2742,
 was subtracted from both roll-angle images to create two separate subtraction
images for each filter.  The images were then geometrically corrected for the
slight optical distortion of the NICMOS camera 2 at the coronagraphic focus.  We used the x-direction pixel
scale of 0\farcs07595/pixel and the y-direction pixel scale of 0\farcs07542/pixel to create an image with pixels that have the y-direction plate scale
in both directions.  The geometrically corrected images were rotated to a common celestial orientation
using the  rotation centers given by the flight software in the raw data file
headers and artifacts such as diffraction spikes were masked.   Figure \ref{fig:f1} shows the resulting PSF subtracted and roll-combined images of the TW Hya disk taken in the medium band filters.

Additional observations of TW Hya with NICMOS were performed as a part of GTO programs 7226 and 7223 with HST \citep{Weinberger:2002}.  The observations in F110W and F160W 
were recovered from the archive, reduced in the same manner as the medium band data, and median combined.  Archival 
PSF reference stars for
the images were
subtracted from the target observations.  The F110W reference was $\tau^1$ Eridani, while the F160W reference was Gl~879.  We followed the procedure of \citet{Weinberger:2002} for scaling the PSFs, but followed the above procedure for 
subtraction. 

TW Hya has been observed with STIS using a combination of wedge coronagraphic
images as well as spatially resolved coronagraphic spectroscopy in the visible, including point source spectroscopy of TW~Hya using the G750L grating and the 52$\times$0.2 slit.  Additional STIS spectroscopy was also obtained for the PSF reference HIP 32939.  The reduction for the data is detailed in \citet{Roberge:2005}.  

\section{Results}
Multi-wavelength, spatially-resolved, scattered-light observations provide a powerful look into both the TW Hya disk's morphology and the wavelength and stellocentric-distance dependent scattering efficiency of its grains.  The spectroscopy of TW~Hya additionally places some constraints on its spectral type.  In this Section we measure the surface brightness profiles, azimuthal asymmetries, and the scattering efficiency of the dust in the TW Hya disk.  We also determine TW Hya's stellar spectral type using its STIS optical/NIR spectrum.

\subsection{Radial Surface Brightness Profiles in the Medium Band Filters}

The surface brightness profiles of TW Hya reveal structure on top of a smooth decrease in surface brightness with radial distance; this structure deviates from what one would expect from a simple flared disk.
We investigate the radial surface brightness profiles in our medium band NICMOS data and compare the behavior in these wavelengths to that seen in the shorter wavelength data.

Figure \ref{fig:f3} shows the F171M through F222M azimuthally averaged surface brightness profiles.  In all of them we see the characteristic shift in behavior between $\sim$80-130~AU that is seen in the visible, 1.1, and 1.6\micron, namely a change in the slope of the surface brightness.  This feature is now seen in all the wavelengths of light in which TW~Hya is observed, indicating that this is a feature caused by some change in morphology rather than a compositional feature localized in wavelength.  We overplot simple flared disk model surface brightness profiles consistent with those used to describe the TW~Hya disk in the sub-mm \citep{andrews12}, which clearly do not match the behavior of the surface brightness profiles for TW~Hya.  

\begin{figure}
\plotone{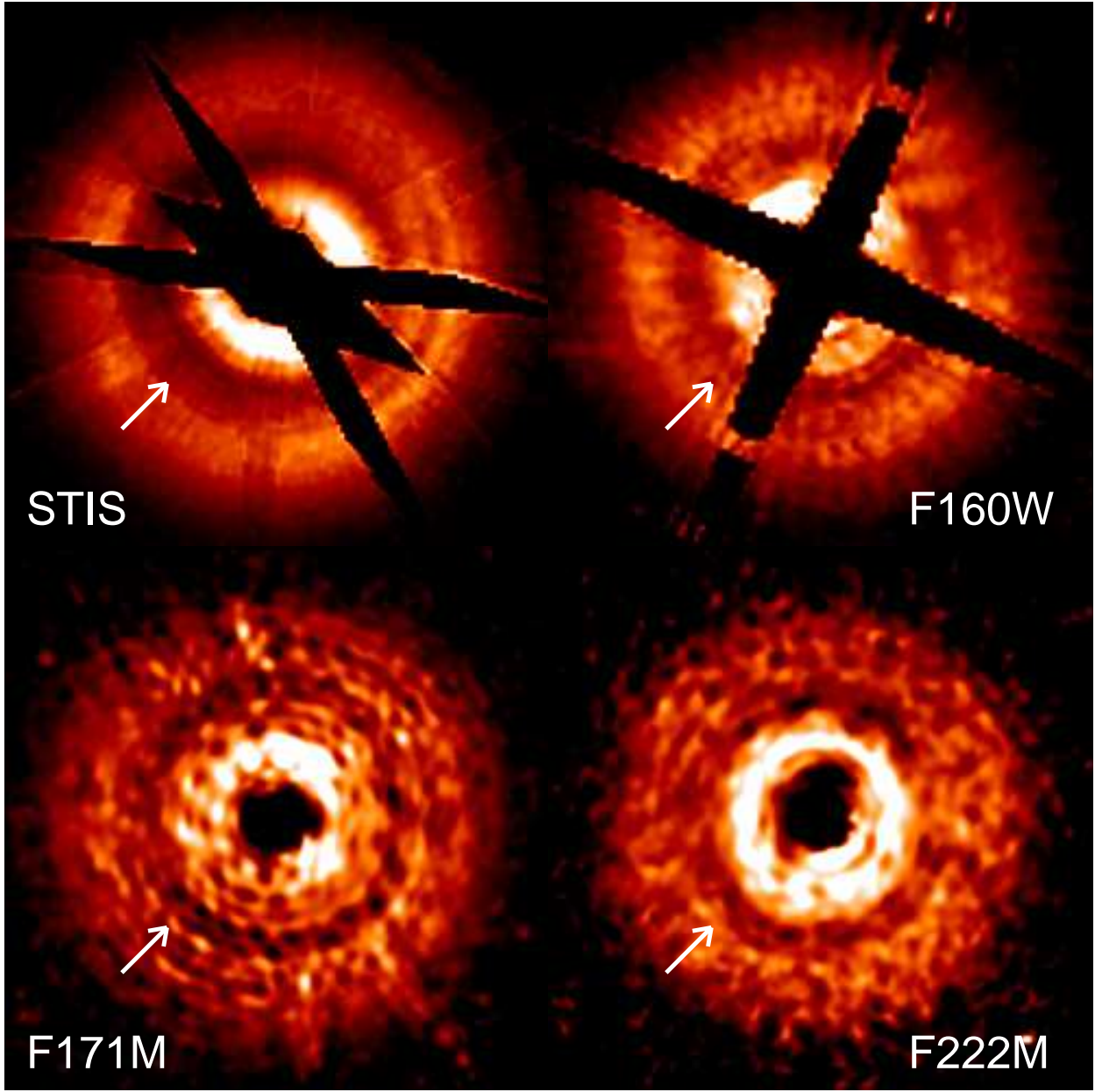}
\caption{\label{fig:f2} Surface brightness STIS, F160W, F171M, and F222M images where the
  surface brightness at each pixel is multiplied by $R^2$, where $R$ is the distance from the central star.  This scaling highlights structure in the disk and shows the sharp cut-off in disk surface brightness exterior to $\sim$150~AU and the apparent depression in surface brightness at 80~AU from the star (location of arrows).  The F171M and F222M images have been smoothed with a 3-pixel FWHM Gaussian filter to highlight the depression in surface brightness. Black regions within the disk are masked due either to the wedge coronagraph in the STIS data, or diffraction spikes in the F160W data.}
\end{figure}

To highlight the structure in the disk, we scaled each pixel of the disk images by $R^2$,
where $R$ is the distance from the central star.  Figure \ref{fig:f2} shows the STIS,
F160W, F171M, and F222M scaled images, which clearly shows a depression in the disk
structure at about 80~AU, coincident with the slope changes in the surface brightness
profiles.  The structure is most pronounced in the higher spatial resolution STIS images,
but is still visible at longer wavelengths.  Additionally, the STIS image shows a possible arc structure exterior to the gap at a PA of $\sim$270$^\circ$.

The depression, or deficit of surface brightness relative to the surrounding disk material can be caused by several factors, including a drastic change in dust opacity at $\sim$80~AU, a sudden change in turbulence of the disk, shadowing due to structure just interior to 80~AU, or the presence of non-axisymmetric structures like spiral arms \citep{weinberger99,Clampin,fukagawa04}.  It could also be caused by a protoplanet accreting material and opening a gap structure within the disk \citep{bate,bryden99}.  Current planet formation models have difficulty forming large planets far from the central star.  However examples of such systems, like HR~8799 \citep{marois}, suggest that large bodies can form around stars at distances of several tens of AU.

In \S \ref{s:gap} we investigate how a physical gap in material might change the observed surface profiles and use models to determine limits on the depth of the gap as constrained from our surface brightness profiles as a function of wavelength.  From these limits we make a preliminary estimate to the limiting mass of a protoplanet capable of opening such a gap in the TW~Hya disk.

\subsection{Azimuthal Surface Brightness Asymmetries}
\label{s:phase}
\citet{Roberge:2005} measured the surface brightness as a function of azimuth for the STIS image of TW Hya's disk and found it possessed a significant asymmetry between 65-140~AU consistent
with the asymmetry caused by a disk inclined to the line of sight with forward scattering dust grains.  The magnitude and position of the asymmetry corresponded to a measured PA of the brightness maximum of 233.6$^\circ\pm$4$^\circ$ and an inferred inclination of 15$^{+8.7\circ}_{-6.4}$ under the assumption of a Henyey-Greenstein scattering phase function for the grains with an asymmetry parameter $g$ of
0.5.  They found no significant asymmetry for larger radii.  Recent CO line data places the PA of the disk at 150-155$^\circ$ \citep[][]{andrews12,rosenfeld12}, consistent with a PA for the side of the disk closest to Earth being either at 60$^\circ$ or 240$^\circ$.  \citep{rosenfeld12} determined that they could reproduce this apparent brightness asymmetry with a warped inner disk at 4~AU with an inclination of 8$^\circ$, rather than 6-7$^\circ$ as inferred by CO line data for the outer disk.  This particular configuration also helps to explain larger than expected CO line velocities in the inner disk regions.

In this paper we re-analyze the STIS image as well as the other NICMOS NIR images to
search for similar asymmetries.  For each passband, we took the final subtracted images
and constructed photometric apertures in the shape of concentric annuli as a function of
increasing radius with equal spacings of 0\farcs357~(20~AU), to improve S/N.  Each annulus was then
further equally subdivided into azimuthal bins with width 20$^\circ$.  For each image a
total of 6 annuli were used, from 0\farcs88~(47~AU) to 2\farcs5~(134~AU).  For the STIS
and F110W images, there was enough signal-to-noise to extend out to 4\farcs4~(236~AU).  We
therefore had a measure of the surface brightness distribution between 50-130~AU for all
passbands, and extending to 230~AU for STIS and F110W.

In the NICMOS images, certain areas were completely masked either by the 
coronagraphic spot or diffraction spikes.  For the STIS image, masked areas included diffraction spikes
and the coronagraphic wedge.  For apertures where masked pixels comprised more than 2/3 of
the aperture, no surface brightness was calculated.  Each annulus was then
scaled to the median of the brightest annulus to provide equal weight to the
surface brightness in any one azimuthal bin at each radius.  The final azimuthal surface
brightness distribution was then constructed by taking a median of the first
six annuli for all images and another distribution was constructed for the
STIS and F110W image of the disk from the final four annuli.  For the
uncertainty at each point we took the quadrature sum of calculated photometric uncertainty and
the standard deviation of the azimuthal points at that angular position.

\begin{figure}
\plotone{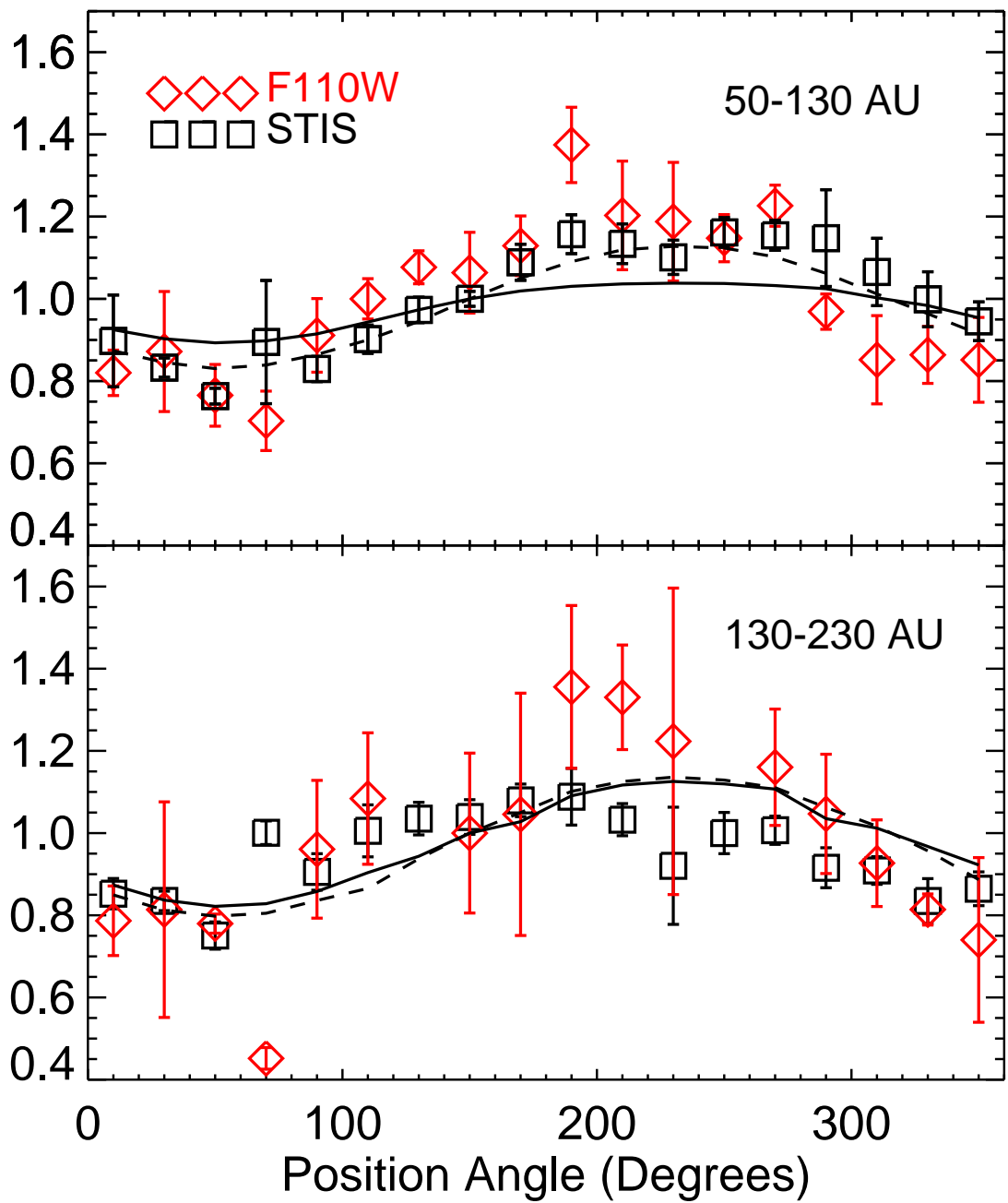}
\caption{\label{fig:f4} Azimuthal brightness profiles for the STIS and F110W images between
50-130~AU (top) and 130-210~AU (bottom). Each curve was normalized by the median surface brightness.  In each panel we compare the brightness asymmetry with that of our best fitting model disk in \S \ref{sec:bestfit} with inclinations of 7$^\circ$ (solid line) and 7$^\circ$ with a $g$ parameter of 0.5 (dashed line).  Both models assume that the PA of the disk is aligned with the side of the disk closest to the observer.  For the isotropic model the disk PA is 53.6$^\circ$, while the forward scattering model has a PA of 233.6$^\circ$.}
\end{figure}

Figure \ref{fig:f4} shows the STIS and F110W results.  Between 50 and 130~AU the asymmetry
is similar in the two passbands.  Beyond 130~AU the azimuthal profile is of a lower
signal-to-noise but roughly matches what is seen at shorter distances, contrary to what
was reported in \citet{Roberge:2005}.  The distance of 130~AU corresponds to a break in
the radial suface brightness of the disk.  For a comparison, we also overplot our best
fitting scattered light models of the disk, described in more detail in \S
\ref{sec:bestfit}, assuming an inclination of 7$^\circ$, with Henyey-Greenstein $g$ parameters of 0 and 0.5.  Even with no forward scattering, flared disks can present a brightness asymmetry when circular apertures are used due to a ``foreshortening'' effect.  Even though the side of the disk closest to Earth is slightly brighter due to the line-of-sight angle to the observer from the disk surface, this occurs at a smaller stellocentric angular separation than that expected for a non-flared disk.  In this scenario, the side of the TW Hya disk closest to the observer would be to the NE.  For STIS and F110W, a model with moderate forward scattering can also reproduce the azimuthal brightness asymmetry, if the semi-minor axis of the TW Hya disk facing the observer is toward the SW.

\begin{figure}
\plotone{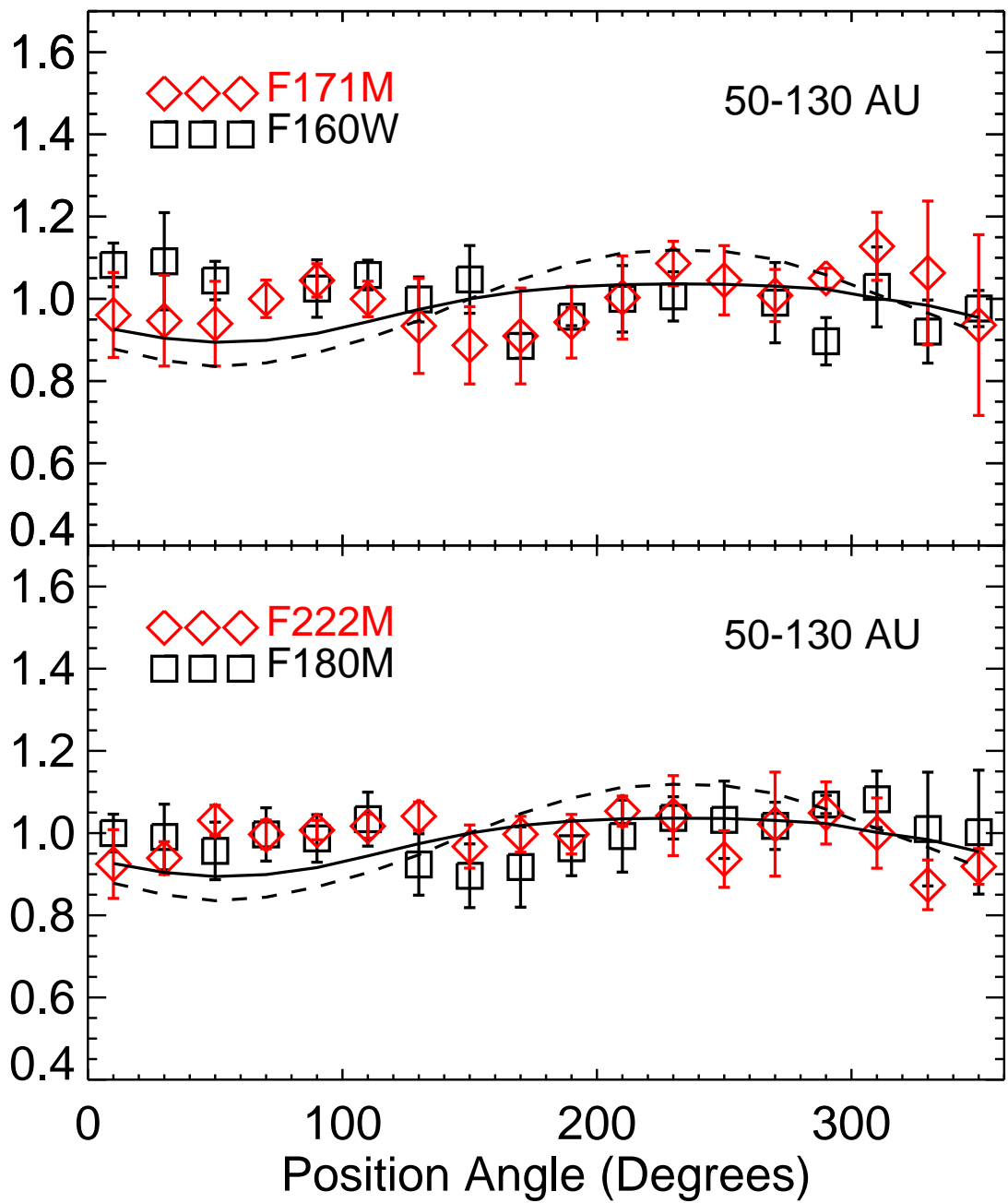}
\caption{\label{fig:f5} Azimuthal brightness profiles for the F160W and F171M images between
50-130~AU (top) and the F180M and F222M images between 50-130~AU (bottom). Each curve was normalized by the median surface brightness.  In each panel we compare the brightness asymmetry with that of our best fitting model disk in \S \ref{sec:bestfit} with inclinations of 7$^\circ$ (solid line) and 7$^\circ$ with a $g$ parameter of 0.5(dashed line).}
\end{figure}

Figure \ref{fig:f5} shows the F160W, F171M, F180M, and F222M azimuthal brightness profiles
with the same models overplotted.  We neglect the F204M data because it has lower
signal-to-noise.  For each of these four images, no significant azimuthal asymmetry is
revealed.  Whatever is the cause of the asymmetry at shorter wavelengths, it is not
detected with significance in the longer wavelength observations.  We discuss the
potential causes of this in Sections \ref{sec:bestfit} and \ref{s:conc}.

\subsection{Scattered Light Spectrum} 

The scattered light spectrum of the disk around TW Hya is a combination of the input
stellar spectrum and the intrinsic reflectivity of the disk.  We removed the stellar spectrum
of TW Hya by dividing the measured surface brightness by the flux density of the star in
each filter.  We divided the STIS spectrum of the disk by the point source spectrum of the
central star.

TW Hya did not have direct photometric measurements in the broadband HST filters (F110W, F160W, and
STIS CCD) and has a complicated spectral energy distribution in both the optical and
near-IR (See \S \ref{s:spectype}).  To find TW Hya's broad-band flux densities, we
bootstrapped from its flux relative to the PSF reference star. Although the PSF does not
have exactly the same spectrum as TW Hya, it is anchored to TW Hya by literature
photometric measurements of both stars at V, J, and H-bands and our measurement at
F171M. To interpolate to STIS, F110W, F160W, we fit stellar atmosphere models of a range
of effective temperatures and gravities to the broad-band photometry of the PSF, scaled to
TW Hya, and propagated the systematic model uncertainty into the TW Hya photometry. We
also compared our flux densities in the near-IR to the spectrum of \citet{vacca},
and found general agreement within the uncertainties.

\begin{figure}
\plotone{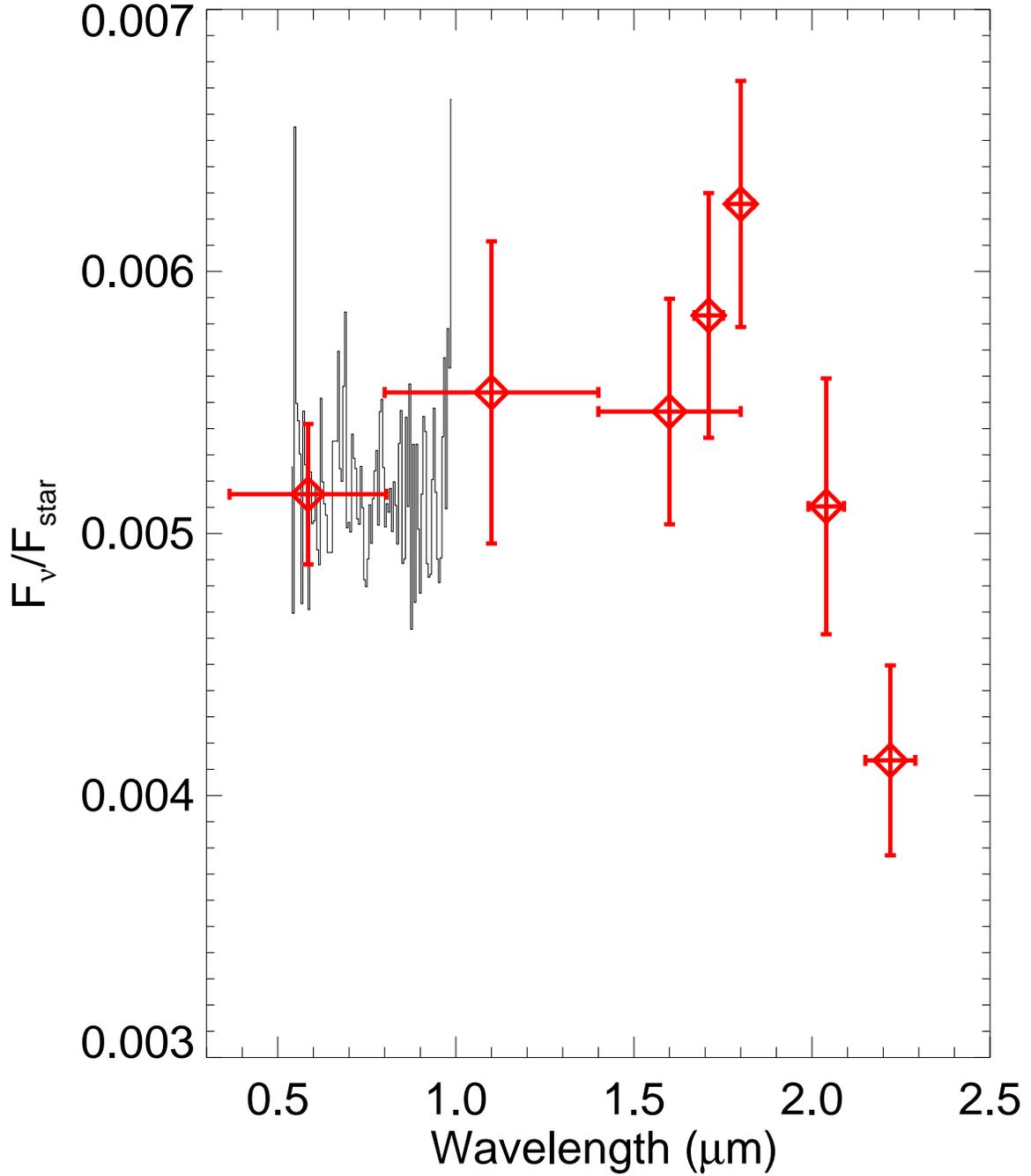}
\caption{\label{fig:f6} Total spectral reflectance of the TW~Hya disk from
  0.5-2.22\micron, including an average spectrum from \citet{Roberge:2005} normalized to
  the STIS disk photometry (black line).}
\end{figure}

Figure \ref{fig:f6} shows the total reflectance spectrum of the disk from
0.5-2.22\micron\ averaged over radial distances of 50-215~AU, the extent to which we
detected the disk in the medium band filters. Also shown is the visible light spectrum of
the disk over the same distances, normalized to the STIS photometric point, from
\citet{Roberge:2005}.  The overall spectrum is relatively neutral between 0.5-1.6\micron,
becoming slightly blue at longer wavelengths.  No absorption bands are seen within this
broadband spectrum.
  
\subsection{The Spectral Type of TW~Hya and its Inferred Mass and Accretion Rate}
\label{s:spectype}
Most previous determinations of TW~Hya's spectral type, and hence its age and mass, have
been based on optical spectral diagnostics \citep{webb,alencar02,yang:2005}.  The
consensus value of the optical spectral type is K7 or an effective temperature of
$\sim$4000~K and a mass of 0.5-0.8~M$_\odot$.  Recent near-IR spectroscopy of TW~Hya,
however, implied a much later spectral type of M2.5 \citep{vacca}, which corresponds to a
lower mass and younger age.

\begin{figure}
\plotone{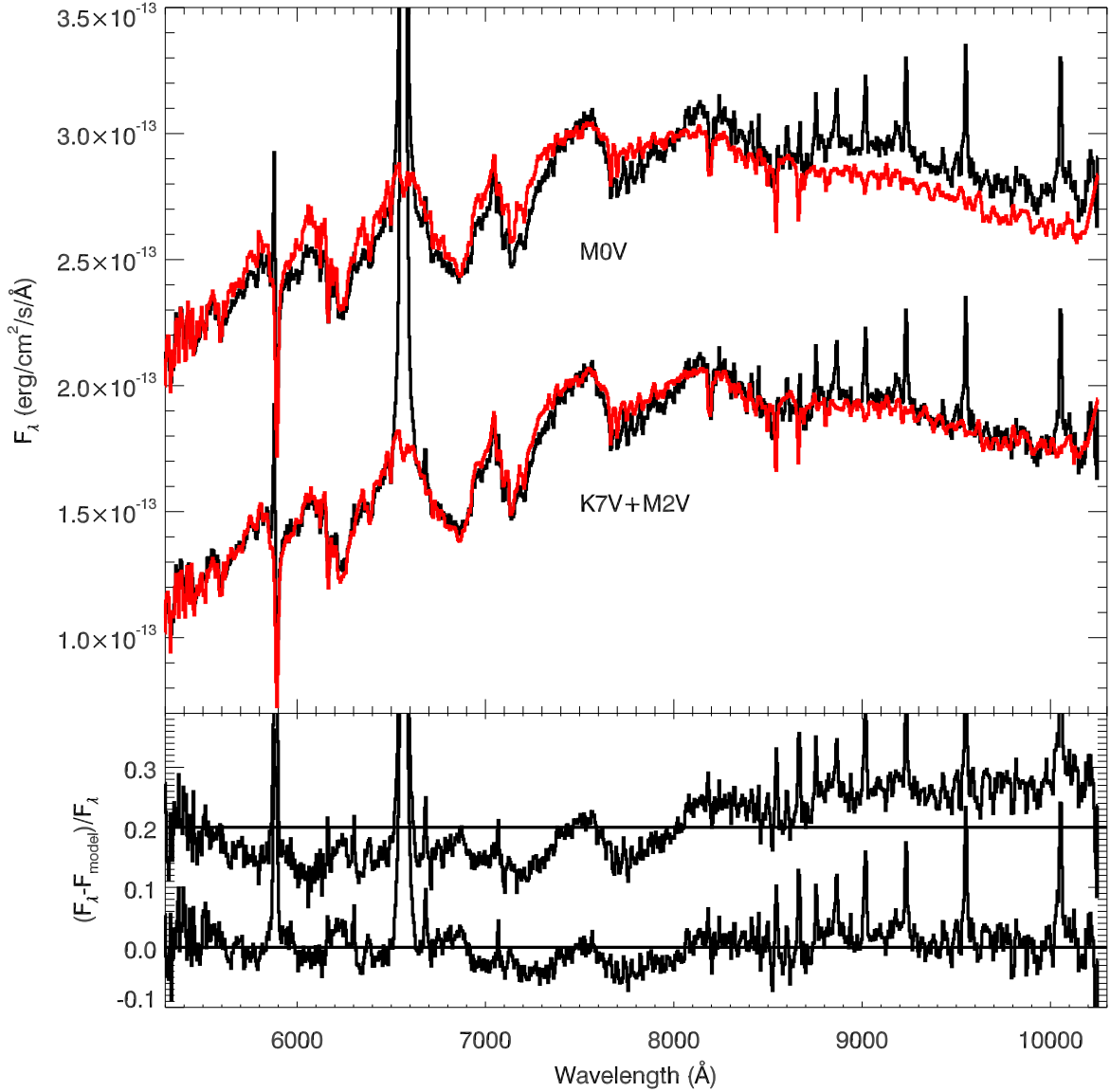}
\caption{\label{fig:spec1} Comparison between the TW~Hya STIS spectrum and two comparison spectra, one from the M0V star  GJ~825 (offset for clarity) and the other a combination of the K7V star HR~8086 and the M2V star HD 95735.  The bottom panel shows the fractional residuals, again with the M0V fit offset upwards for clarity.  The M0V fit, while decent, cannot simultaneously match both the blue and red ends of the spectrum as the two component fit can.}
\end{figure}

We investigated this discrepancy by finding our own spectral type for TW Hya using the
broad wavelength coverage of our G750L spectrum (5240-10266\AA) taken 17 July 2002. We
compared our spectrum to those of K7-M2 stars in the STIS Next Generation Spectral Library
\citep{heap1,heap3,gregg1,gregg2}, whose spectra also incorporate STIS G750L data. We
calculate a reduced $\chi_\nu^2$ value from a comparison of the spectra at each wavelength
 with each of the six comparison stars listed in
Table 2 as models, while excluding data in the H and Ca emission lines.  The reduced $\chi_\nu$ value
 was determined against the 961 wavelengths from the spectrum, for a total of 960 (958) degrees of 
 freedom for a one- (two-)component model respectively.  For each model
star, we determined a mean T$_{eff}$ and uncertainty from the literature, computed over
the range of published T$_{eff}$s.  The closest match ($\chi_\nu^2$=105) was to GJ 825, a
M0V dwarf with an effective temperature of 3730 $\pm$ 160~K (Figure \ref{fig:spec1}).  The
fit is rather poor at both wavelength extremes -- TW Hya is dimmer in the blue and
brighter in the red.

\citet{vacca} suggested that TW~Hya is in fact a later type star with a hot spot.  An
alternative explanation is that TW Hya is a hotter star with substantial cool spot, such
as the one imaged near the magnetic poles and seen in radial velocity measurements
\citep{huelamo08,donati11,boisse12}.  In order to test these hypotheses, we further fit our
spectrum with combinations of two comparison stars of different spectral types.  A
significantly better ($\chi_\nu^2$=35), though not perfect, fit is obtained with a
combination of a K7 (45\% of the flux) and M2 (55\% of the flux) star.  Our K7 comparison
was HR~8086 (T$_{eff}$=3990$\pm$115~K) and our M2 comparison was HD 95735
(T$_{eff}$=3600$\pm$130~K).  This combination provides a good fit to both the optical and
NIR spectrum of TW~Hya.  One limitation to our method is the lack of multiple comparisons
for a finer temperature constraint, as well as disagreement on the T$_{eff}$ of our
comparison stars in the literature.

We find the effective radius of TW Hya by combining the known distances to the comparison
stars and to TW~Hya with interferometrically measured \citep{vanbelle} or model-inferred
\citep{takeda07} stellar radii of the comparison stars. We find a radius of
0.84$\pm$0.2~R$_\odot$ for the M0 comparison and 1.08$\pm$0.15~R$_\odot$ for the K7+M2
comparison, roughly in agreement with the findings of \citet{vacca}.  We assume the
uncertainties in the radius are driven by the uncertainty in the parallax.

\begin{figure}
\plotone{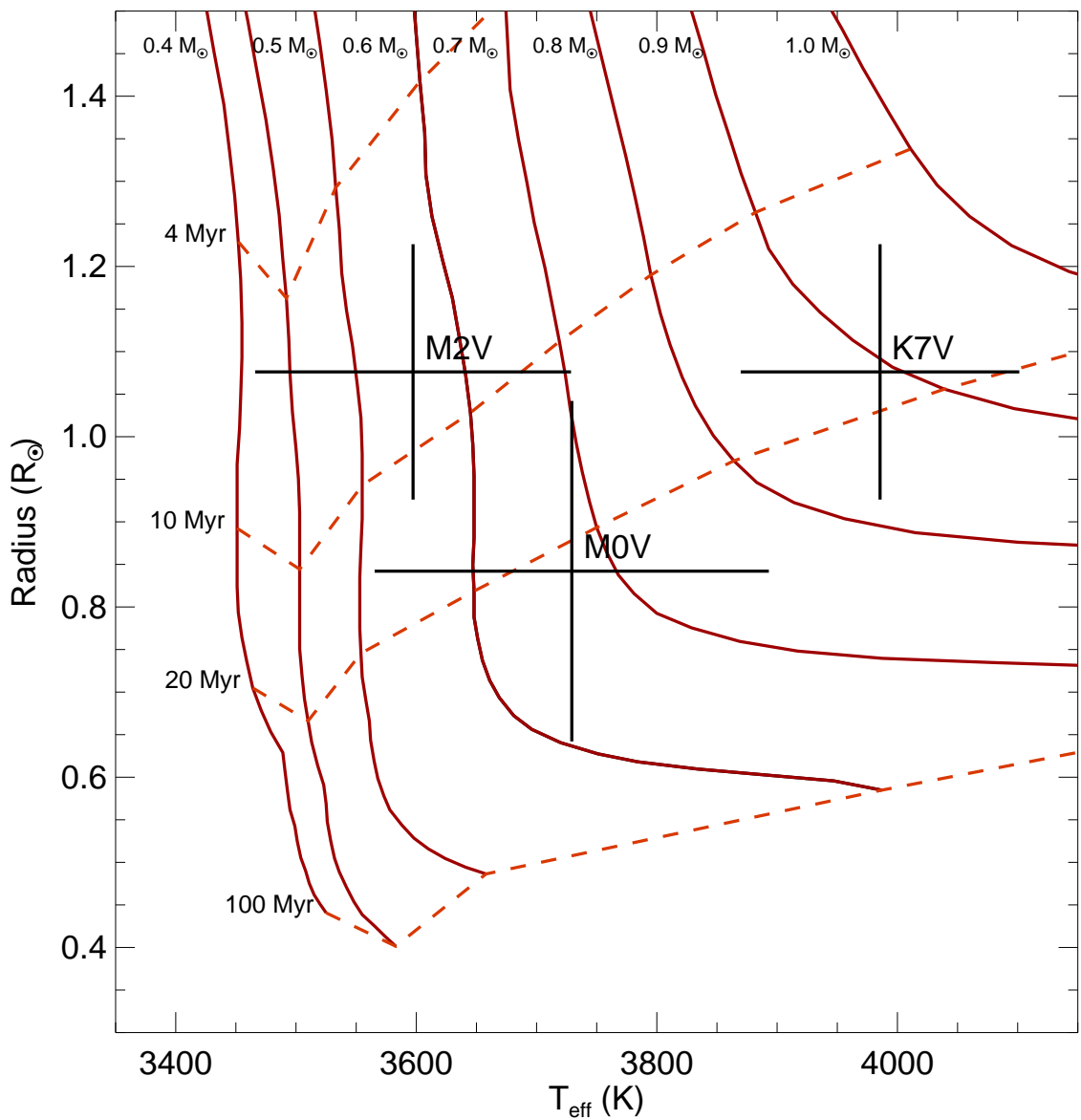}
\caption{\label{fig:ff} Comparison of isochrones (dashed lines) and evolutionary tracks (solid lines) from \citet{baraffe98} to inferred T$_{eff}$ and radii for TW~Hya from our fitting to the STIS spectrum of TW~Hya.  Our best fit to a single component is to an M0V comparison star, while our best fit to a two component model requires either a cooler M2V star with a hotter accretion spot component that is similar to a K7V spectral type, or a hotter K7V star with a cooler M2V spot component.  The most consistent match to the TWA moving group's estimated age is that of a cool M2V star with a mass of 0.55~M$_\odot$ and an age of 8~Myr.}
\end{figure}

This complex situation involving multiple temperature components makes an inference of a
mass, T$_{eff}$, and luminosity for TW~Hya problematic. However, a comparison of our
inferred temperature and radius with evolutionary models provides some insight into
whether we have a low mass star with accretion hot spots and a higher mass star with large
cool spots.  We overplot the inferred T$_{eff}$ and radii from our spectral templates in
Figure \ref{fig:ff} with respect to isochrones for stellar masses of 0.4-0.8~M$_\odot$
from \citet{baraffe98} (inspection of other isochrones give similar results within the
uncertainties).  If the coolest component represents the central star, then the inferred
radius and T$_{eff}$ are consistent with a 0.55$\pm$0.15~M$_\odot$ object at an age of
$\sim$8~Myr, in line with other age and mass estimates of TW Hya.  If the hot component
dominates, however, the age would be closer to 20~Myr.  Similarly, the intermediate case
of a slightly higher T$_{eff}$ object comparable to an M0 spectral type implies an older
star (t$\sim$20-30~Myr).  

We conclude that the best interpretation of stellar spectrum is of a cooler star
(T$_{eff}\sim$3600) with a mass of 0.55$\pm$0.15~M$_\odot$ and an age of 8$\pm$4~Myr.
There exists, therefore, a significantly warmed atmosphere due to accretion over portions
of the star's surface.  An estimation of the accretion luminosity can come from assuming
the K7 component as the SED of the warmed photosphere due to accretion (additional emission may arise at shorter wavelengths due to the hotter shock itself).  Under the assumption of blackbody emission, if we subtract off the underlying luminosity
of the M2V component, the luminosity coming from accretion is estimated to be
0.03~$L_\odot$, corresponding to an areal covering fraction of 10\%.  Assuming $L=GM\dot{M}/R$, this implies accretion rates for TW~Hya
of 2$\times$10$^{-9}$~M$_\odot$ yr$^{-1}$, in line with other estimates of TW~Hya's
accretion rate \citep{batalha02}, which range from
4$\times$10$^{-10}$-5$\times$10$^{-9}$~M$_\odot$/yr.

Sub-mm measurements of the Keplerian rotation of gas in the TW~Hya disk in principle provide an independent measure of the stellar mass.  Given the near face-on inclination for the disk, even very precise measurements of Keplerian rotation in TW Hya provide a similar constraint to mass as to what we have obtained through our STIS spectroscopy:  while CO measurements of TW Hya's disk constrain the inclination to 6-7$^\circ$ and accuracies to within a degree, that still allows for a wide range of stellar mass, since the quantity $M^{1/2}\sin{i}$ is measured.  If the inferred inclination is 6$^\circ$ and the mass of TW Hya is best fit by 0.6$M_\odot$ \citep{hughes11}, an uncertainty of 1 degree would correspond to masses that ranged from 0.85$M_\odot$ to 0.44$M_\odot$   \citep{qi04,hughes11} 
Most of the prior
SED and sub-mm spectral cube modellng have assumed M$_*$=0.6 M$_\odot$
\citep{Calvet:2002,Qi:2008}, while other investigators have assumed masses as high as M$_*$=0.8 M$_\odot$ \citep{donati11,andrews12}.  In \S \ref{s:model}, we select our best fitting mass of 0.55 $M_{\odot}$ and determine that mass does not significantly impact most parameters of interest within the disk.

\section{Modeling the TW Hya Scattered Light Disk}
\label{s:model}
In this Section we provide a model for the observed spectrum and morphology of the disk. This requires a model of the structure of the disk, discussed in \S\ref{s:diskstruct}. We infer that the depression we observe between 50-130~AU
is caused by a ``gap'' in the disk. The origin of this feature will not be constrained by these observations, but a likely candidate is a planetary object that has succeeded in opening a gap. The limits to such a candidate are discussed in \S\ref{s:planetmass}. However, other possibilities exist, which we outline in \S\ref{s:conc}.

\subsection{Initial Disk Structure}
\label{s:diskstruct}

The disk models are generated as described in \citet{HJC_gaps}.  The details
of the radiative transfer modeling is also described in \citet{paper1,paper2}
and \citet{HJC_model}. Although we assume that the overall disk structure is
axisymmetric, the calculation of the radiative transfer is done in 3D to
include the curvature of the disk both in vertical height above the midplane
($z$) and in the azimuthal direction ($\phi$).

The structure of the unperturbed planet-less disk is generated in a two-step
procedure. In the first stop, we calculate a locally plane-parallel two-dimensional model for
the entire disk.  We use the same formalism developed by \citet{calvet} and
\citet{vertstruct,dalessio2}, with some simplifying assumptions.  We calculate
the disk from 0.25 to 256 AU, with radial bins spaced by factors of
$\sqrt{2}$.  At each radius, we assume that the disk is locally
plane-parallel.  This quickly generates an estimate for the radial and
vertical temperature of the disk and the surface density profile that we can
then refine in step (2).

Then, in step (2), we remove the assumption of a local vertical plane parallel
surface, but keep the vertically integrated surface density profile fixed.  We
take a radially limited slice of this disk and refine its structure, this time
taking the full three-dimensional (3D) curvature of the disk into account. We
calculate radiative transfer in 3D as a numerical integration over the surface
of the disk.

In both steps, we iteratively calculate the vertical density and temperature
structure of the disk including radiative transfer and under the assumption of
hydrostatic equilibrium.  The main heating sources are stellar irradiation and
viscous dissipation.  The opacities used are the same as those used for
calculating the disk brightness, although to speed up the calculations, the
opacities are averaged over the stellar and disk thermal emission.  To
represent the optical depth of the disk to stellar light, we average the
wavelength-dependent opacity over the Planck spectrum evaluated at $T_{\rm
  eff}$ and we use the Rosseland mean opacity to represent the optical depth
of the disk to its own radiation.
  
We base our disk structure modeling on our inferred stellar parameters
from Section \S \ref{s:spectype}.  We assume a stellar mass of
0.55~M$_\odot$, radius of 1.08~R$_\odot$, and total 
luminosity of 0.208 $L_\odot$.  The spectrum of the star 
is composed of two blackbodies, one at 3990 K and the other 
at 3600 K, contributing 55\% and 45\%, respectively, to the 
total luminosity.  The opacities of the dust to stellar irradiation 
are calculated using this spectrum.  

Given the uncertainties in mass and T$_{eff}$ described in \ref{s:spectype}, we
consider their impact on our models.  Decreasing the stellar mass while keeping
the total luminosity of the star fixed has the effect of making the disk puffier, due to
the lower gravity.  The disk is then able to intercept more light, making it appear
brighter.  Disk models differing only by the stellar parameters have the same
basic shape and spectral behavior but change overall brightness by 5-10\%.  Changing the
stellar parameters also mainly impacts the inferred radial width of the gap, but not the
depth.

The disk parameters are those derived by \citep{Calvet:2002} 
from the SED of TW Hya.  
For the accretion rate, we assume a value of 
$1\times 10^{-9}\, M_{\sun}\mbox{yr}^{-1}$ (see \S \ref{s:spectype}).  
We assume that the dust is well-mixed with the gas
and constant throughout the disk, with a dust-to-gas 
ratio of 0.0138 and a grain-size distribution of 
$n(a)\propto a^{-3.5}$.  
\citet{Calvet:2002} and \citet{Wilner:2005} found that large maximum grain sizes
($a_{\mbox{\scriptsize max}}$)
give good fits to the SED of TW Hya at long wavelengths
($\lambda \gtrsim 1$ cm), but 
there is a degeneracy between maximum grain size 
and viscosity parameter ($\alpha$) in their models, 
where the viscosity is given by 
$\nu = \alpha c_s^2/\Omega_{\phi}$ where $c_s$ is the thermal sound speed, 
and $\Omega_{\phi}$ is the Keplerian orbital angular velocity.  
Good fits to the SED of TW Hya are obtained with
$a_{\mbox{\scriptsize max}} = 1\,\mbox{mm}, 1\,\mbox{cm},$ or 
$10\,\mbox{cm}$.  This is because large grains are effectively 
invisible to optical and infrared wavelengths.  
The $\alpha$ parameter must be adjusted as the 
maximum grain size changes, because the overall mass of the 
disk increases with increasing grain size, assuming a fixed 
gas-to-dust ratio; however, the stellar mass accretion rate must 
be kept fixed.   

Given that $a_{\rm max}$, $\alpha$, and $\dot{M}$ all 
have the primary effect of scaling the disk mass up or down 
and that the degeneracies inherently give limited constraints 
on these parameters, we assume for simplicity in all our disk models that 
$\dot{M} = 10^{-9}\, M_{\sun}\mbox{yr}^{-1}$,
$a_{\mbox{\scriptsize max}} = 1\,\mbox{cm}$, and
$\alpha = 5\times10^{-4}$.  
The effect of increasing the 
accretion rate would be to increase the total surface 
density of the disk, which could be offset by adjusting 
the viscosity parameter $\alpha$ downward.  
Our assumed parameters give a total disk mass within the simulation boundaries between 27 and 211 AU 
of 0.074 $\Msun$. 
This assumed disk mass is comparable to estimates of the disk mass from 
Herschel HD observations \citep{berginnature}.

The wavelength dependent opacities are calculated using a 
Mie scattering code, miex, that includes distributions of particle sizes and 
can account for large dust particles \citep{wolf04}.
The disk models rely on mean opacities calculated from 
these wavelength-dependent opacities.  
The Rosseland mean opacity evaluated at 100 K is used to 
represent the overall disk opacity to its own emission 
and the Planck averaged opacity evaluated at 100 K is 
used for calculating the thermal emission from the disk.  
The Planck averaged opacity evaluated over the 
stellar spectrum is used to represent the opacity 
of the disk to stellar irradiation.  
We model the star as a two-temperature 
blackbody, with 55\% of the luminosity from a 
3990 K component, and 45\% of the luminosity from a 3600 K 
component.  
For a model with a given minimum grain size, we calculate 
the entire disk model from the initial conditions 
in order to be completely self-consistent.  

\subsection{Scattered Light}
\label{sec:model}

The scattered light image of a disk is modeled as in
\citet{2009HJC} and \citet{HJC_gaps}. 
The scattering surface of the disk is defined to be
where the optical depth from the star at a given frequency $\nu$ is
$\tau_{\nu} = 2/3$. The brightness
at the scattering surface of the disk, including multiple scattering, is
\begin{equation}\label{eq:multscat}
I^{\mbox{\scriptsize{scatt}}}_{\nu} =
\frac{\omega_{\nu}\mu\/R_*^2\/B_{\nu}(T_*)}{4r^2(\mu+\cos \eta)}
\left\{
1 + \frac{\omega_{\nu}}{1-g^2\mu^2} 
  \left[\frac{(2+3\mu)(\mu+\cos\eta)}{(1+2g/3)(1+g\cos\eta)} - 3\mu^2
\right]
\right\}.
\end{equation}
where $\omega_{\nu}$ is the albedo,
$g=\sqrt{3(1-\omega_{\nu})}$, 
$\mu$ is the cosine of the angle of incidence to the scattering surface,
$B_{\nu}$ is the stellar brightness,
$r$ is the total distance to the star,
and $\eta$ is the angle between the line of sight to the observer and
the normal to the scattering surface. 
The observed brightness of a star at distance from the observer $d$ is
\begin{equation}
F_{\nu,\mbox{\scriptsize{obs}}} = \pi B_{\nu} \left(\frac{R_*}{d}\right)^2
\end{equation}
so we can express the surface brightness in scattered light
in units of the apparent brightness of the star per square arcsecond:
\begin{equation}\label{eq:Iscatt}
I^{\mbox{\scriptsize{scatt}}}_{\nu} =
\frac{\omega_\nu \mu (1+f_{\rm mult})}{4\pi(\mu+\cos{\eta})}
\left(\frac{d}{\mbox{pc}}\right)^2
\left(\frac{r^2+z_s^2}{\mbox{AU}^2}\right)^{-1}
\left(\frac{F_{\nu,\mbox{\scriptsize{obs}}}}{\mbox{asec}^2}\right)
\end{equation}
where
\begin{equation}
f_{\rm mult} = 
\frac{\omega_{\nu}}{1-g^2\mu^2} 
  \left[\frac{(2+3\mu)(\mu+\cos\eta)}{(1+2g/3)(1+g\cos\eta)} - 3\mu^2
\right]
\end{equation}
represents the fractional increase in brightness 
caused by multiple scattering.

Finally, we need to choose a composition and size distribution (especially minimum grain
size) of the dust to calculate the final scattered light that is emitted from the disk.
We use the wavelength-dependent opacities calculated using Mie theory, as described above,
to determine the scattered brightness.  Many compositions and possible grain sizes are
available, and it is not immediately clear whether a single composition is preferred for
optically thick disks, or whether each individual disk has its own properties.  The
neutral color of TW~Hya in the optical compared to the quite red colors of HD~100546,
suggest that a range of compositions and size distributions exist amongst disks.

\subsection{Gap and Truncation}
\label{s:gap}
The gap in the disk is modeled as an ad hoc axisymmetric density
perturbation parameterized by a Gaussian with adjustable
width $w$ and depth $d$ centered at 80 AU.
To model the truncation of the disk, we introduce an exponential
cutoff at a knee $k$, as appears in the self-similar 
solution for an accretion disk as derived in \citet{1998HCGD}.  
If $\Sigma_0(r)$ is
the unperturbed disk surface density, then the new surface
density profile is
\begin{equation}
\Sigma(r) = \Sigma_0(r) \{1-d\exp[-(r- 80\mbox{ AU})^2/(2w^2)] \}
\exp[ -(r/k) ].
\end{equation}

Since the primary heating source for the disk is stellar irradiation,
the effects of shadowing and illumination must be accounted for in
determining the vertical structure of the disk. 
The three-dimensional density and temperature structure of the
perturbed disk are calculated iteratively according to \citet{HJC_model}
and \citet{HJC_gaps} 
That is, the illumination at the surface of the disk is determined
by ray-tracing and the disk temperatures are calculated accordingly. 
Once the disk temperatures are determined, the vertical density
profile of the disk is iteratively
recalculated assuming hydrostatic equilibrium
as above, keeping the vertically integrated surface density profile
constant.

\subsection{The Predictive Power of Scattered Light Observations of Optically Thick Disks}

How predictive can models of optically thick disks be for composition and size
distribution?  While we discuss below (\S\ref{sec:bestfit}) the best fit to our
measurements of TW Hya's disk, here we consider the uniqueness of composition within such
fits.  We consider a fixed disk geometry by taking a single disk model from the parameter space we explored for TW~Hya--a disk with a gap located at 80~AU with width $w$=30~AU and depth $d$=0.3, a
truncation knee at $k$=100~AU, and a maximum grain size of 1~cm.  We then made six
independent models of the disk structure that vary composition between pure water ice and
astronomical silicates \citep{warren84,laor93}, and vary minimum grain sizes between
5$\times10^{-3}$,1, and 10\micron.  The dust is well-mixed with the gas so that the dust
density is proportional to the gas density. We calculated albedo and opacity to find the
surface brightness of scattered light as a function of wavelength.

\begin{figure}
\plotone{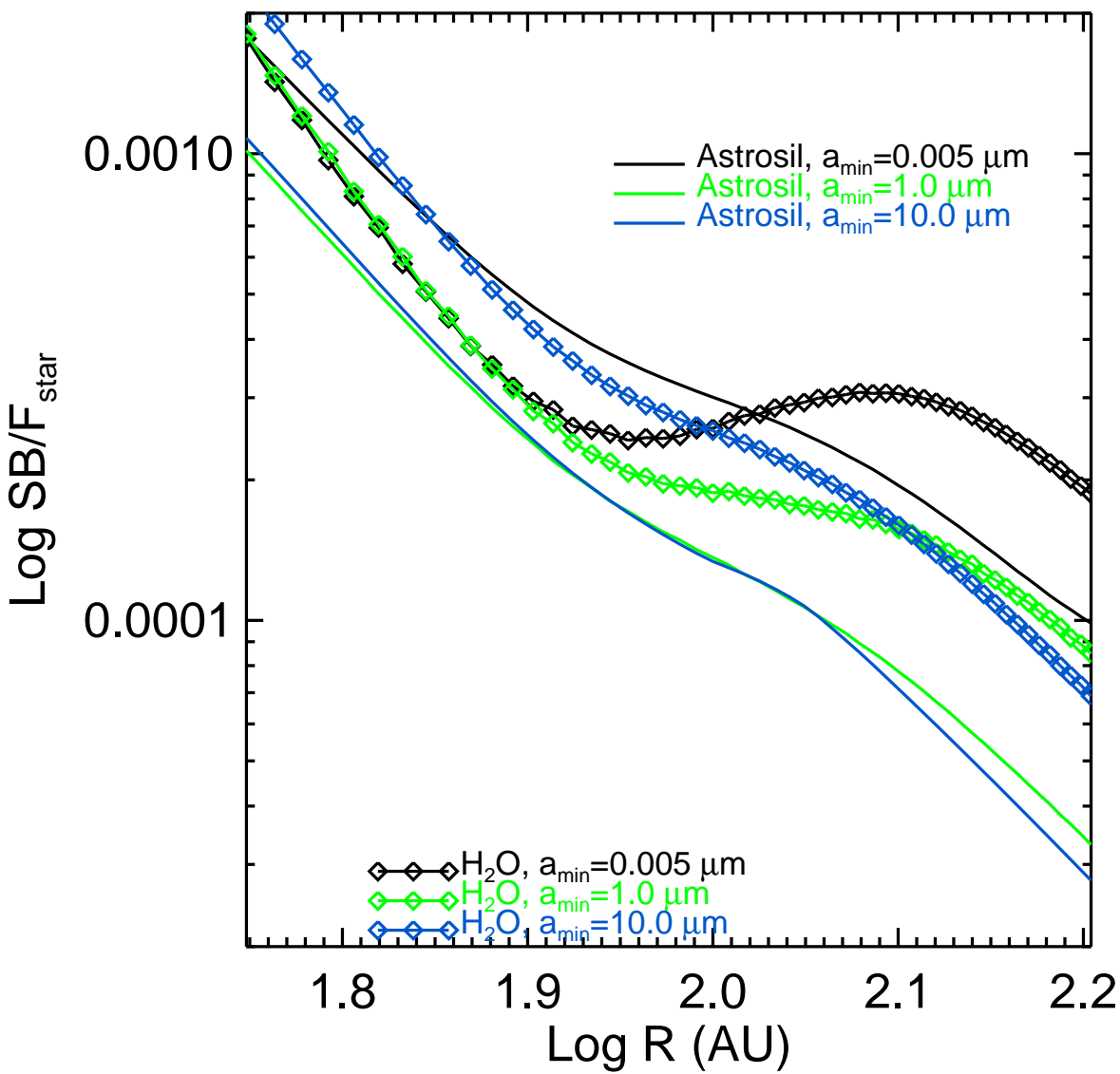}
\caption{\label{fig:f7} Examination of the affect of grain composition and minimum size on the surface brightness profile of a TW Hya-like disk.  The disk structure is held constant, and pure water ice (squares) and pure astronomical silicate (solid lines) compositions compared.   Silicates tend to make the disk brighter because they have a higher scale height }
\end{figure}

Figure \ref{fig:f7} shows a comparison between the six predicted STIS surface brightness
profiles for the different models.  Differing grain sizes and compositions indeed affect
the resulting surface brightness profiles of a given disk structure.  If the composition
of the disk is incorrect, the structure of the disk can be incorrectly interpreted.

\begin{figure}
\plotone{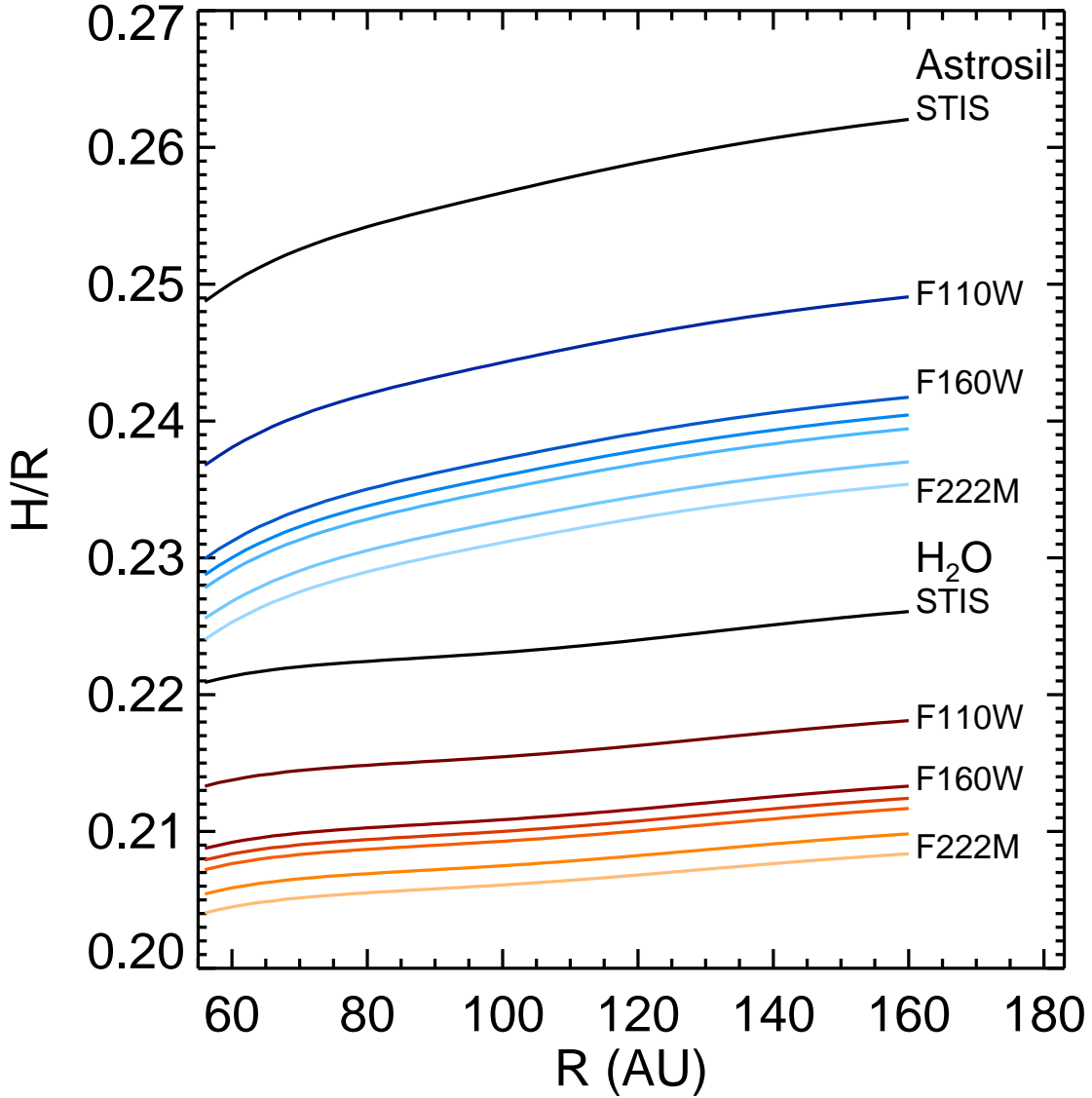}
\caption{\label{fig:f8}A comparison of scale height, H/R, for astronomical silicates (blue lines) and water ice (red lines) with $a_{min}$=0.005\micron, where H is the $\tau=2/3$ optical depth surface at each wavelength.  As expected, longer wavelength filters probe deeper within the disk surface, though the specific geometry of that surface changes with composition.}
\end{figure}

\begin{figure}
\plotone{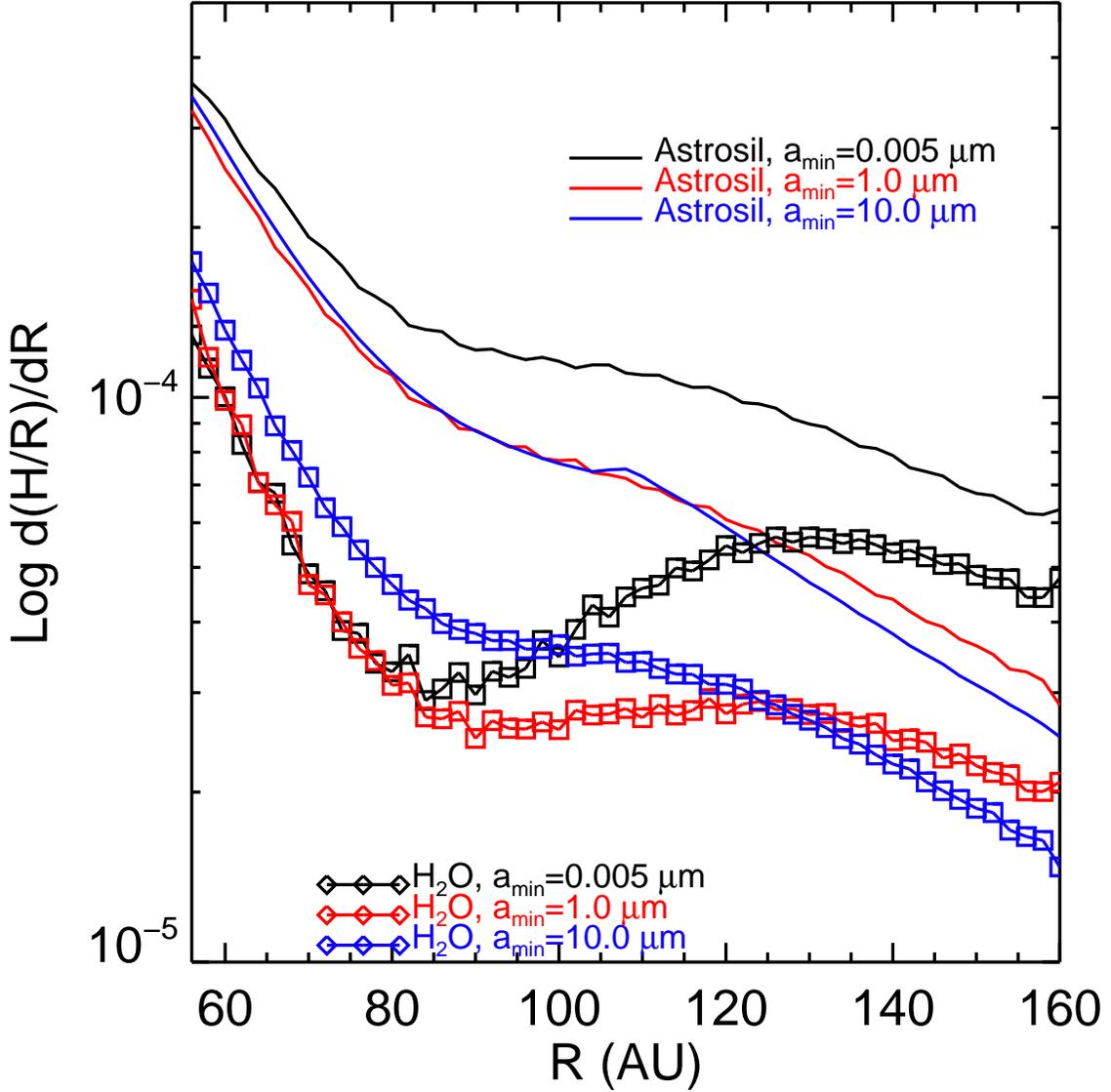}
\caption{\label{fig:f9} Comparison of the slope of the disk surface, i.e.  d(H/R)/dR, for
  pure water ice and astronomical silicate compositions of differing $a_{min}$.  The
  derivative of the disk surface goes into the calculation of the angle of incidence of
  starlight on the disk and therefore the total scattered light as discussed in \S \ref{sec:model}.}
\end{figure}

In general, pure water ice disks are dimmer than those that possess pure silicates.  The reason for this difference is primarily due to the higher opacity of a pure silicate disk, demonstrated in Figure \ref{fig:f8}.  In this Figure we have plotted the height of the $\tau_{\nu}=2/3$ surface ($H$) divided by the radius ($R$) for varying wavelengths.  
From Eq.~(\ref{eq:Iscatt}), the brightness of the disk is proportional 
to the angle of incidence $\mu$ at the surface, 
\begin{equation}
\mu \approx \frac{d}{dR}\left(\frac{H}{R}\right) - \frac{H}{R}
= R \frac{d}{dR}\left(\frac{H}{R}\right)  
\end{equation}
Note that $H$ represents the wavelength-dependent penetration depth 
of stellar photons, not a thermal scale height.  
If $H\propto R^{\beta}$, then $\mu=(\beta-1)H/R$.  
Therefore, the brightness scales roughly with $H/R$.  
The $H$ surface occurs higher up in the disk for silicates, 
resulting in a higher surface brightness.  
The different structure of the surface brightness profiles in Figure
\ref{fig:f7} is also striking, and is clarified by looking at
$d(H/R)/dR$ for the different models (See Figure \ref{fig:f9}).

\begin{figure}
\plotone{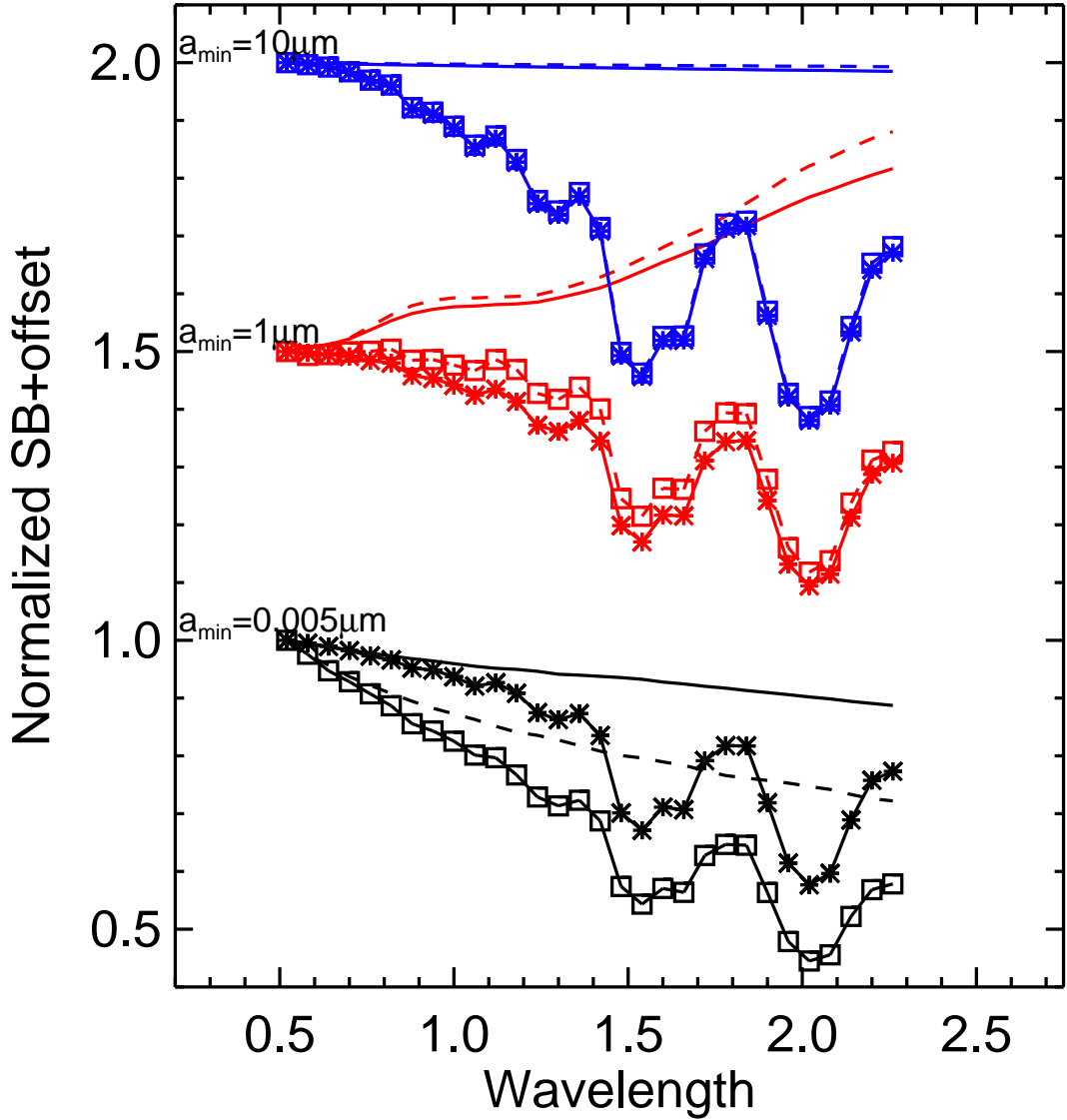}
\caption{\label{fig:f10} Comparison of TW Hya-like disk reflectance spectra for different pure water ice (squares and asterisks) and pure astronomical silicate (solid and dashed lines) compositions.  Squares and solid lines represent the interior of the disk at 56~AU, while asterisks and dashed lines represent the outer part of the disk at 140~AU,}
\end{figure}

Water ice has strong absorption features at 1.5 and 2.0 \micron\, which are probed by our
F160W/F171M and F204M observations.  Given the $\sim$10\% accuracy of the disk photometry
measurements, we can in principle detect absorption features that have a depth of
$\sim$15-30\%.  Figure \ref{fig:f10} shows the resulting reflectance spectra for each of
our models.  Silicates show no absorption lines in the visible to near-IR, but are mostly
neutral or red over this wavelength range for minimum grain sizes $>$1\micron.  Water ice
can show noticeable features, and disks with pure water ice can show detectable features
at 20-40\% depth.

These tests demonstrate that composition can play an important role in the observed
scattered light spectrum and surface brightness profiles of a disk.  For our final model
of TW Hya, an exhaustive search of compositions, grain sizes, and structures is
computationally prohibitive and beyond the scope of this paper. 

\subsection{Model Parameters}
\label{sec:bestfit}
To fit the observations of TW Hya's disk, we used dust opacities and scattering
efficiencies for a composition with the same relative abundances as used to model the SED
(\citep{dalessio3}): 29.6\% organics, 40.4\% water ice, 24.5\% astronomical silicates, and
5.5\% troilite.  We test whether such a composition also fits the scattered light
observations.  We ran a suite of models, varying the following disk parameters as follows:
\begin{eqnarray}
a_{\mbox{\scriptsize min}} & \in& 
        \{0.005, 0.5, 5\} \mbox{ microns} \label{avar} \\
w &\in& \{5, 10, 20, 30\} \mbox{ AU}   \label{wvar} \\
d &\in& \{0.0, 0.3, 0.5\} \label{dvar} \\
k &\in& \{60, 80, 100, 120, 150\} \mbox{ AU}  \label{kvar}
\end{eqnarray}

The surface brightness profiles are fit simultaneously 
spatially and across the 7 wavelengths for which we have photometric data.  
To account for uncertainties in the overall normalization such as might 
arise from uncertainty in our chosen parameters, we allow 
the overall brightness to vary by a constant factor 
and calculate the reduced $\chi^2$ value for each model 
with respect to the wavelength dependent surface brightness profiles from $56$ to $160$ AU\@.
The center of the gap is fixed at 80 AU.  

The parameters for the best fits as measured by the minimum reduced $\chi^2_{\nu}$ from the fits above and are
tabulated in Table \ref{tab:fitparams}, where we found that a disk gap of 30\% depth and 20~AU width are preferred.  
The profiles are well fit by a truncation knee of 100~AU.  We found that we could fit the observed
brightness of the disk only with small a$_{\rm min}$ (0.005$\micron$), but if we were
willing to scale the overall brightness, we could equally fit (similar $\chi^2$) the
disk with a$_{\rm min}$=0.5$\micron$. However, larger grain sizes failed to reproduce the
neutral-to-blue color observed over large portions of the disk, resulting instead in a
redder color.  Since the overall flux density of the disk is an observed quantity, we
favor those models that naturally reproduced the observed surface brightness.

A few of our assumptions could impact the expected brightness of the disk.  For example,
the luminosity of the star would differ by $\sim$20\% if we instead used the luminosity
determined by \citet{vacca}, rather than that by \citet{webb} or our own determination.
Since the surface brightness of TW~Hya's disk due to scattered light depends on the square
of the distance, uncertainties of up to 23\% are possible, given the uncertainties in
TW~Hya's parallax.  A different distance would also imply revised stellar parameters.  A
higher or lower mass of the central star will result in a puffier (brighter) or more
vertically compact (dimmer) disk, respectively.  

We have also assumed that the dust isotropically scatters light.  If the dust were instead
forward scattering by a moderate degree, such as with a Henyey-Greenstein asymmetry
parameter of 0.5, the approximate scattering angle of TW~Hya would result in a dimming of
the disk of $\approx$25\%, based on our modeling.  Due to flaring in the disk, our isotropic models predict a
decrease in flux along the semi-minor axis of the disk pointing to the observer.
Conversely, forward scattering dust, even in a flared disk, will produce a brightness peak
along the semi-minor axis of the disk tilted toward the observer.  The true orientation of
the disk is degenerate between these two possibilities, so we orient our models such that
the brightness maximum is aligned with the observed P.A. in the STIS and F110W data.

Closer to the star, the models with isotropic scattering do not completely reproduce the
azimuthal brightness distribution, while the forward scattering model is more consistent.
Our models suggest an inclination of $\sim$7$^\circ$ and forward scattering dust for the
disk.  Forward scattering grains in the outer disk are marginally favored as both the STIS
and F110W data still show a trough and peak structure in the azimuthal surface brightness
profile--while the S/N is lower, the expectation from our models is a larger brightness
asymmetry at larger distances, which is not favored by the data.  This is either
indicative of a decrease in propensity to forward scatter or lack of forward scattering
grains in the outer disk.

\begin{figure}
\plotone{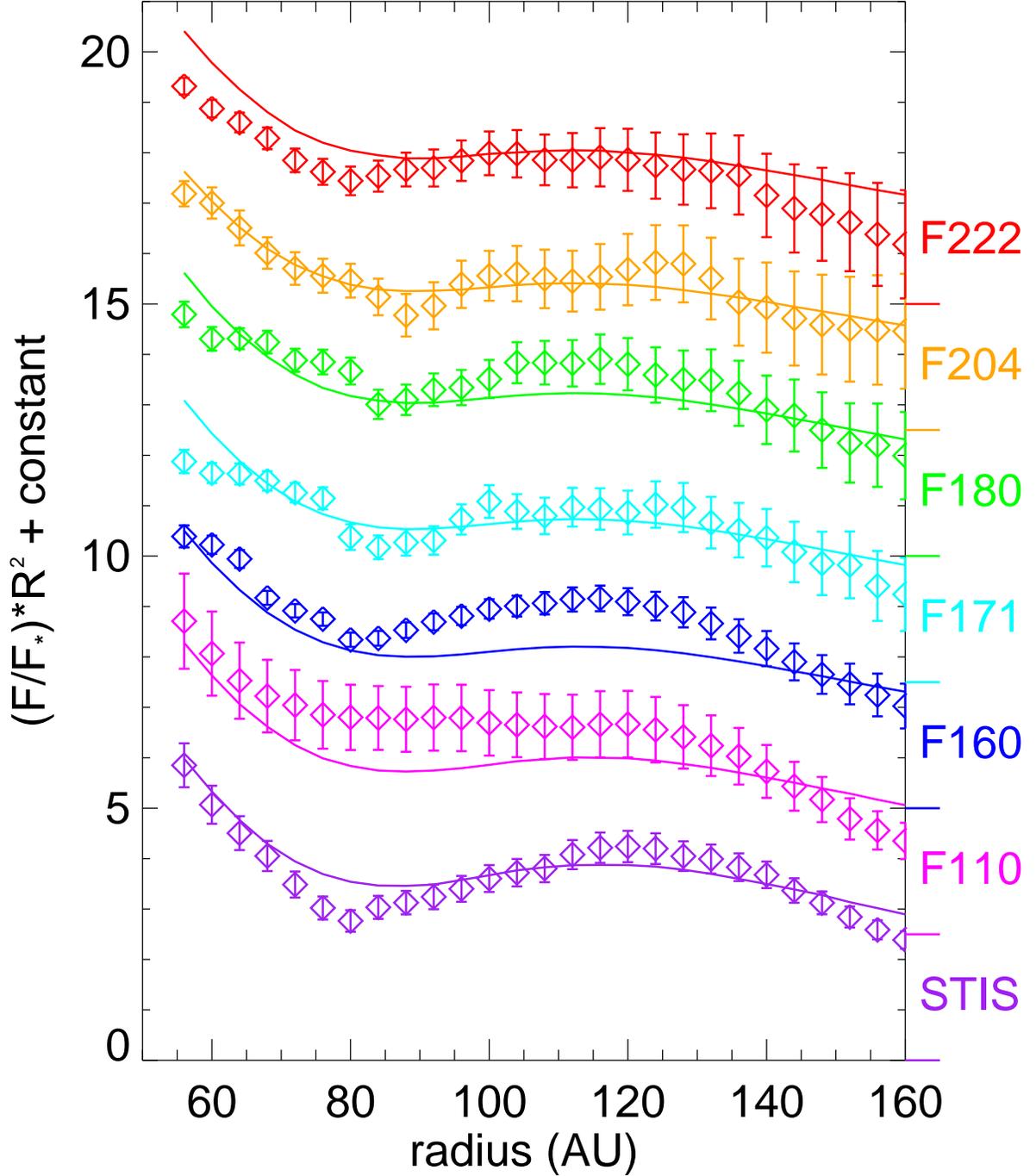}
\caption{\label{fig:f11}
Comparison of normalized radial profiles of the disk with the 
best fitting model for an assumed stellar mass of 0.55~$\Msun$.  
The fit parameters are tabulated in Table \ref{tab:fitparams}.  
The brightness profiles have been multiplied by $R^2$ to 
emphasize the gap shape, and are separated by constant factors of 2 
for clarity. 
}
\end{figure}

Figures \ref{fig:f11} and \ref{fig:f12} show the radial data
compared to the best fit model for all of the observations and radial
spectral cuts of the disk.  In general the model fits the observed
spectrum quite well, with the exception of the region around 80~AU, 
the position of the gap.  
This is also where the model fails to reproduce the observed surface
brightness profile in some of the wavelengths observed.

In comparison, models with no gap provide poor fits to the wavelength dependent
surface brightness profiles.  
In figure \ref{fig:f3} we show brightness profiles 
of disk models without a gap at 80 AU in comparison to the data 
at the F222M waveband.  The disk models include truncation 
at $k=60$, 100, and 150 AU and have been scaled by 
$1.05$, $0.73$, and $0.57$, respectively, in order to fit the data 
across all wavelengths.  The reduced $\chi^2_{\nu}$ for the 
models are 20.8, 8.18, and 4.31, respectively.  The 
model with truncation at 150 AU is the best fit, but still does 
not provide as good a fit as compared to the model with 
a 30\% gap.

\section{Mass of a Gap-Opening Planet}
\label{s:planetmass}

The best model for the TW Hya disk is one where an axisymmetric  
$30\%$ partial gap is opened 
at 80 AU in the disk.  A possible mechanism for opening such a gap 
is an embedded planet.  The model we have used to fit to the TW Hya disk 
does not include hydrodynamics, 
but rather the gap is imposed in an ad hoc manner  
on the disk structure.  

Numerical hydrodynamic simulations of planets embedded in gas-dominated 
protoplanetary disks by \citet{bate} and \citet{bryden99}
indicate that tidal torques between the planet 
and disk can clear axisymmetric partial gaps.  As described in \citet{HJC_gaps}, the gap size can be correlated to planet mass according to the following arguments.
The mass at which a planet can open a gap depends upon both 
the scale height of the disk and its viscosity.  
The thermal scale height of the disk is 
\begin{equation}\label{thermalscaleheight}
h = \frac{c_s}{v_{\phi}}a
\end{equation}
where $c_s=\sqrt{kT/\mu}$ is the thermal sound speed measured at the midplane, 
$k$ is the Boltzmann constant, 
$T$ is the local disk temperature, 
$\mu$ is the mean molecular weight, 
$v_{\phi}=\sqrt{GM_*/a}$ is the orbital speed of the planet, 
and $G$ is the gravitational constant.  
For a composition dominated by molecular hydrogen, 
$\mu = 2 m_H$.  
For a viscous protoplanetary disk, the criterion for gap-opening
can be expressed as 
\begin{equation}\label{eq:viscgapcrit}
\frac{3}{4}\frac{h}{r_{\mbox{\scriptsize Hill}}} + \frac{50}{q {\cal R}}
\lesssim 1
\end{equation}
\citep{2006CridaMorbidelliMasset}, 
where
the Hill radius of the planet is 
\(r_{\mbox{\scriptsize Hill}} = \left(\frac{m_p}{3M_*}\right)^{1/3} a\), 
 $q=m_p/M_*$ and ${\cal R}\equiv r^2\Omega_P/\nu_v$
is the Reynolds number, and $\nu_v$ is the viscosity,
given by $\nu_v=\alpha c_s h$ for an $\alpha$-disk model. 

\citet{bate} adopt disk parameters of $h/r=0.05$ and ${\cal R}=10^5$ 
for their simulations, giving a gap-opening threshold of 
$q=1.06\times10^{-3}$ using Eq.~(\ref{eq:viscgapcrit}), or 
slightly more than 1 $M_J$.  For comparison, the inviscid gap-opening 
threshold would be 0.4 $M_J$.  
They find that planets with 
$q=3\times10^{-4}$ and $1\times10^{-4}$ clear gaps of 
90\% and 50\%, respectively, and a planet with $q=3\times10^{-5}$ 
creates a nearly negligible gap.  
Thus, a planet that clears 
a gap of 30\% would be between 0.03 and 0.1 of the viscous gap-opening 
threshold.  

For TW Hya, our best-fit disk model has $h/r=0.081$ and 
$\alpha=5\times10^{-4}$, giving 
${\cal R}=3.1\times10^5$ and a viscous gap opening threshold of 
$q=1.1\times10^{-3}$.  If $M_*=0.55\,M_{\sun}$, this is 
197 $M_{\earth}$. This implies that if an embedded planet is the 
cause of the gap we have modeled, it must be between 
$6-20\,M_{\earth}$.  However, there is some degeneracy 
with regard to $\alpha$, as discussed in \S\ref{s:diskstruct}.  
A model with maximum grain size of 1 mm and $\alpha=1\times10{-3}$ 
gives a disk with similar properties to the best-fit model.  
This larger value of $\alpha$ gives a lower Reynolds number, 
requiring a more massive planet to open a gap and allowing a 
larger planet to hide in the disk.  If we say that 
$\alpha<1\times10^{-3}$, 
we find a gap opening threshold of $q<1.5\times10^{-3}$.  
Thus, a more conservative upper limit on the mass of a 
planet at 80 AU in the TW Hya disk is 28 $M_{\earth}$.

\citet{2006CridaMorbidelliMasset} estimate the half-width of the 
gap to be $\sim 2 r_{\rm Hill}$, based on the region where 
gas streamlines form horseshoe orbits.  
If we take $q=10^{-4}$ and $a=80$ AU, then the gap width would be 5.1 AU\@.  
However, this width does not correspond to the widths of the 
Gaussian-shaped gaps modeled in this work, since the gap shapes modeled in 
\citep{2006CridaMorbidelliMasset} are flat-bottomed.  
A Gaussian fitted to these gap shapes would have widths larger than 
$2 r_{\rm Hill}$.  Thus a $\sim20$ AU gap width, as inferred by our models, 
is only slightly higher than one might expect
for a planet opening a gap.

\section{Discussion}
\label{s:conc}


We have combined several resolved images of the TW~Hya disk in scattered light with
spatially resolved spectroscopy of the disk.  The spectrum of the disk in the visible and
near-IR is featureless, with a broad neutral to blue trend, indicative of sub-micron
grains.  Based on our scattering models, disk grains composed of organics, water ice and
silicates that fit the SED also fit the scattered light, if they are as small as
0.005$\micron$. We have chosen a composition based on what is most likely present in the
TW~Hya disk, but we have not exhaustively explored other materials.  Any material that
scatters neutrally in the visible to near-IR could be suitable to explain the spectral
shape we observe, but would also have to be tested against the SED.  

One implication of our work is that scattered light measurements can effectively constrain
minimum grain size since the smallest dust typically has the largest available surface
area for scattering. TW Hya's disk is very bright and thus requires the presence of small
grains.  By the absence of any strong water ice absorption features, we can also directly
constrain the abundance of water ice on large grains; small grains show absorption
features at $<$10\% the disk scattering continuum for the compositions we considered (see
Figure \ref{fig:f10}), below the level of our disk photometric uncertainties.  A lack of
absorption accords with the expectation that large grains settle below the optical depth
probed by scattered light observations.

For our chosen composition, we can compare our water ice mass abundance to that of
\citet{twhyanat}, whose mass is derived from their Herschel detection of water vapor
caused by UV dissociation at 100-150~AU.  They interpreted the deficit of water vapor to
that predicted by their models as evidence for dust settling of large icy grains to the
disk midplane and calculated that there needed to be $>9\times10^{27}$~g of water ice in
the disk.  Our scattered light images probe deeper in disk scale height than UV and are
also consistent with the idea that the bulk of the ice is on large grains that have
settled toward the midplane.  If we take a disk mass in gas of 1.9$\times10^{-2} \Msun$,
as assumed by \citet{twhyanat}, then we would predict a total ice mass in the disk from our
models of 2$\times10^{29}$~g, based on our mass abundance of 40\% water ice in dust and a
dust-to-gas ratio of 0.0138.  The mass increases by a factor of three if we use the
inferred disk masses from our best fitting model, which has an inferred disk mass closer
to 0.07 $\Msun$.  Assuming that we probe the local mass abundance of water ice relative to
gas at our $\tau=2/3$ surface at the disk, this would imply a relative abundance of
5.6$\times10^{-3}$ at 30~AU above the midplane at 100~AU and 46~AU above the midplane at
150~AU.

Our models must include a 30\% cleared gap at 80 AU and a disk truncation
exterior to 100 AU to match the observed surface brightness profiles. We place
an upper limit on the mass of a planetary companion that could be clearing the
gap of 6--28 M$_{\earth}$.   In the following we discuss potential remaining uncertainties.

\subsection{Remaining Uncertainties}
While our model successfully reproduces the observed properties of the TW~Hya
disk over our wavelength range, there remain some uncertainties that will
require follow-up study.  In particular, we further discuss the nature of the
observed brightness asymmetry for the disk seen at short wavelengths, the more
spatially extended nature of small dust grains in optical light compared to
continuum sub-mm and mm observations, and alternative origins for the observed
gap at 80~AU.

The brightness asymmetry first reported by \citet{Roberge:2005} and further characterized
in \S \ref{s:phase} is strange given the asymmetry is only apparent at wavelengths
$\leq$1.1$\mu$m, whereas our models at any disk inclination would predict asymmetry at
wavelengths $>$1$\micron$ also. A warped inner disk, as suggested by
\citet{Roberge:2005} and \citet{rosenfeld12} also seems unlikely, as that would also be expected to produce
asymmetry at long wavelengths as well.  What could cause a wavelength dependence to the
asymmetry?

For an inclined, optically thick disk with isotropically scattering grains, the near side of the disk appears slightly
brighter because of the lower opacity along the line of sight. However, the
near side also appears foreshortened.  Our azimuthal profiles were taken assuming
a circular disk, which would artificially decrease the surface brightness for position angles corresponding to the 
side of the disk pointing towards the observer.  Therefore, our models do not
uniquely predict the brightness asymmetry.

Forward scattering grains cancel the foreshortening effect by
brightening the near side.  If the disk grains are substantially more forward
scattering at short wavelengths, the apparent brightness asymmetry could
change with wavelength.  However, forward scattering grains would also make
the whole disk look dimmer than observed, since we image it nearly face-on.

TW Hya is known to have variable accretion \citep[e.g.][]{dupree:2012} and
this could drive changes in the scale height of the inner disk. Variability in
the inner disk structure is unlikely as the source of the changing
asymmetry. Although the inner disk could change scale height due to variable
accretion, the F110W, STIS images and STIS spectroscopy show the same
asymmetry but were taken two, and then two more, years apart, respectively,
while the F160W image does not show asymmetry and was taken only three months
before the F110W image.

If portions of the outer disk were shadowed from direct starlight by a
non-axisymmetric bump  or warp in the inner disk, that might be able to produce
asymmetry in the observed scattered light \citep{rosenfeld12}.  The structure would become more
optically thin at longer wavelengths.  The problem is that the
visible-near-infrared extinction would have to be nearly gray to match the observed photometry of the disk.
That only happens from
very large grains, whereas we determine that very small grains are plentiful
in the outer disk.

Alternatively, the lack of asymmetry at longer wavelengths could be the result
of a changing disk structure or grain population with depth, as longer
wavelengths probe deeper in the disk.  The forward scattering grains seen at
wavelengths shortwards of 1.1\micron\ could no longer be present deeper in the
disk to be replaced by a different (perhaps larger) grain population.  Our
models predict where the change-over in grain properties occurs as defined by
the scattering surface between F110W and F160W.  At 80~AU the scattering
surface of the disk is at 19.9~AU above the midplane for 1.1\micron\, while it
is at 19.3~AU for 1.6\micron.  If such a vertical transition exists, it would
be within a very narrow layer of the disk.  If any upper layer structure is
the cause for the azimuthal asymmetries, they must reside only in layers
$>$19.3~AU above the midplane. Such an abrupt change seems unlikely.

A more detailed examination of the effect of forward-scattering grains,
coupled with realistic models of forward scattering from aggregate grains will
be needed to make progress on interpreting the asymmetry.

We next discuss the overall size of the disk.  As has been seen in CO and dust continuum
data at sub-mm and mm wavelengths, the gas in the disk of TW~Hya extends much further than
sub-mm dust emission \citep{andrews12}.  This dichotomy apparently does not extend to
smaller grain-sizes as the scattered light disk is nearly co-spatial with the extent of
the CO gas.  
  
\begin{figure}
\plotone{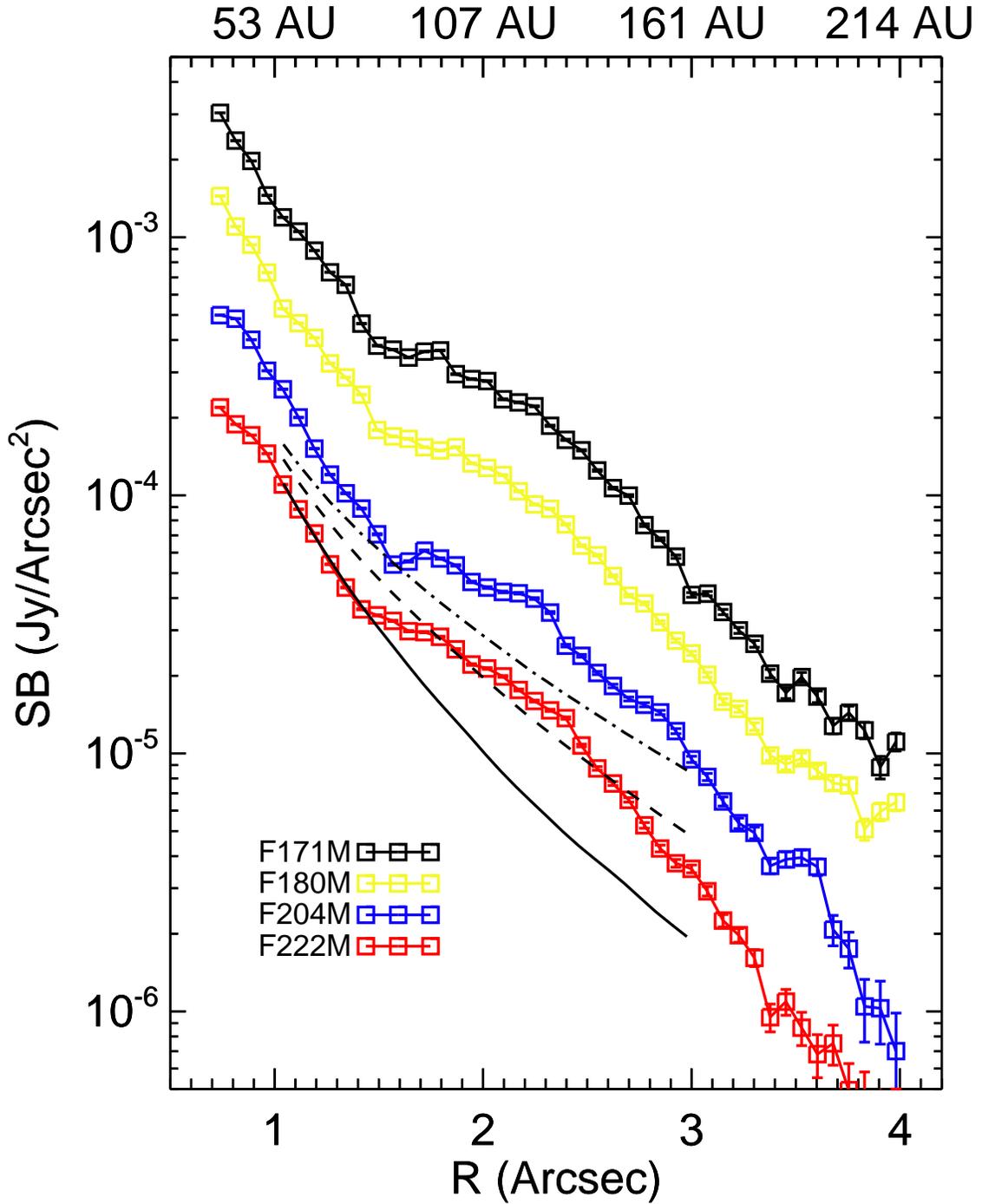}
\caption{\label{fig:f3} Surface brightness profiles of the TW Hya disk at 1.71, 1.80, 2.04, and 2.22 \micron.  The 1.80, 2.04, and 2.22 \micron\ curves are scaled by factors of 0.5, 0.25, and 0.125 respectively.  Overplotted on the F222M data are three gap-less flared disk models with self-similar exponential truncation knees at 60 (solid line), 100 (dashed line), and 150~AU (dash-dotted line; See \S \ref{s:gap}).  None of the gap-less models fit the changes in the SB profile we observe across multiple wavelengths.}
\end{figure}

We find that the best fit model for the scattered light has a 
large truncation knee, at 100 AU, 
although we explored values down to 60 AU.  Smaller values
of $k$ produce steep surface brightness profiles not warranted by our data (See Figure \ref{fig:f3}).
In comparison, radio interferometric observations find a much smaller
truncation radii for similar models.  
Using the similarity solution \citep{LBP1974}:

\begin{equation}
\Sigma(R) = \frac{c_1}{R^{\gamma}} 
\exp\left[-\left( \frac{R}{k} \right)^{2-\gamma} \right].
\end{equation}

\citet{Andrews2012} 
find $k$=35-45 AU for values of $\gamma$ between $-1$ and $1$ 
fit SMA observations at 870 $\mu$m continuum and the CO J3-2 line; 
\citet{Hughes2008} 
find k = 30 AU and $\gamma=0.7$ for SMA observations at
1.3 mm and 870 $\mu$m (345 and 230 GHz) continuum, although 
\citet{Hughes2011} 
find $k=50$ AU and $\gamma=1$ 
is the best fit to the CO J(3-2); and 
\citet{Isella2009} 
find $k = 17.5$ AU and $\gamma=-0.3$
for CARMA observations at 1.3 mm continuum. 
Models by 
\citet{Gorti2011} 
find that a truncation knee
of 35 AU and $\gamma=0.7$
provide good fits to observations of gas emission lines in
TW Hya.

The model fits to radio observations 
cited above predict that the disk density beyond 
$\sim50$ AU should be sharply cut off.  As $k$ and $\gamma$ 
decrease, the more sharply the disk will be truncated.  
However, the scattered light observations clearly indicate 
the presence of disk material beyond 200 AU, so 
we do not expect to find a small truncation knee.  
The scattered light brightness profile of the TW Hya disk suggests that
the disk truncation is at a larger radius than the radio observations
imply. Since scattered light probes only the diffuse surface layer of
the disk while radio observations probe the deeper interior structure
of the disk, the difference in truncation radii may be attributable to
layered disk structure, where the vertical profile of the disk varies
with distance in a way not adequately modeled simply by vertically
integrated surface density profiles.  Conversely, there could be an overall lack of large grains beyond $\sim$60-80~AU, indicative of grain growth frustration beyond this radial distance in the disk.

\begin{figure}
\plotone{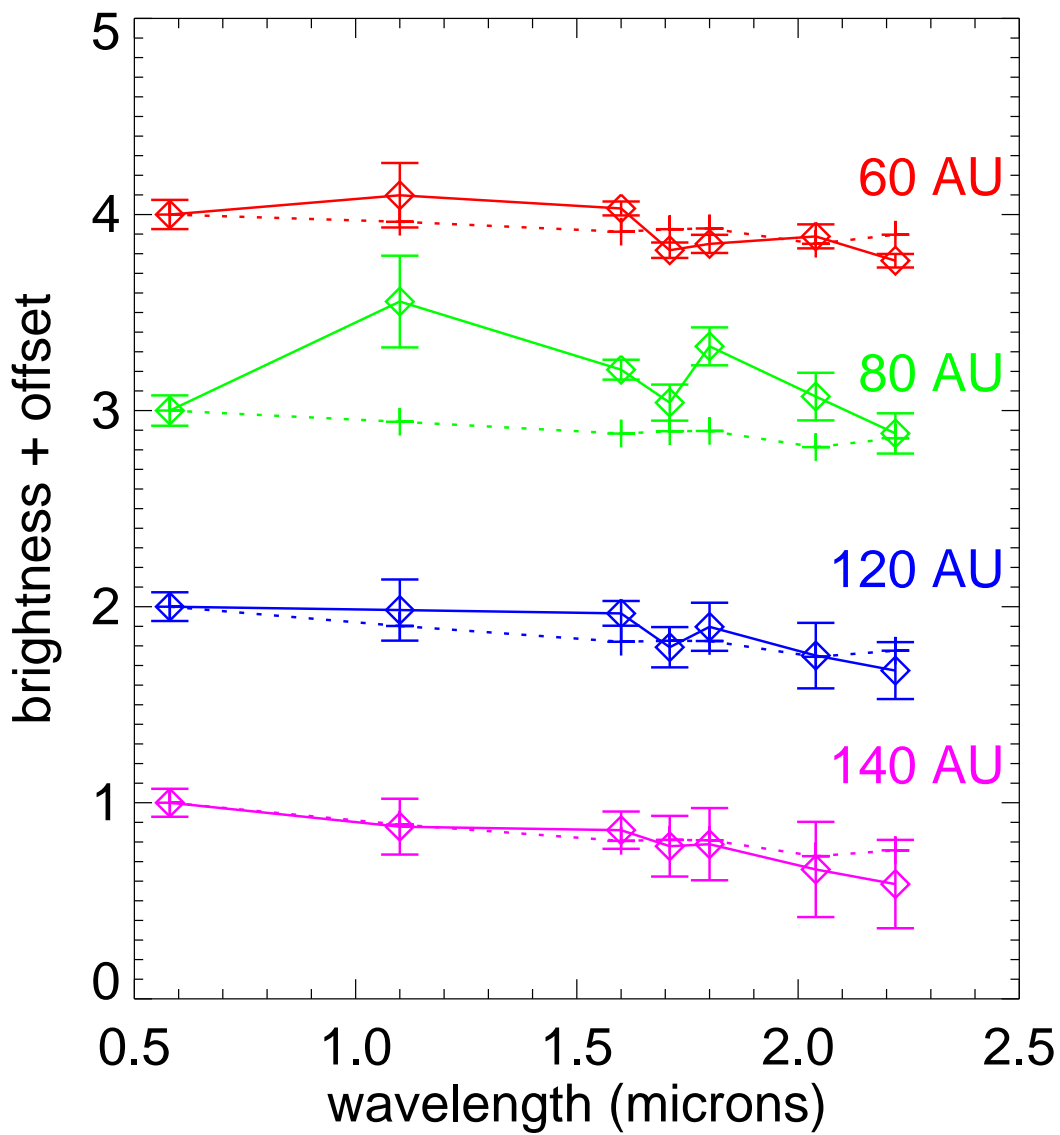}
\caption{\label{fig:f12}Comparison of radial spectra of the disk with model profiles from the best fitting disk models of \S \ref{sec:bestfit}.  The spectra are normalized to the brightness at the STIS wavelength of 0.58 $\mu$m and offset vertically by whole numbers, starting with 3 for 60~AU, 2 for 80~AU, 1 for 120~AU, and 0 for 140~AU.  
The 0.6 (0.4) $\Msun$ model is indicated by dotted (dashed) lines 
and $+$ ($\times$) symbols.  
}
\end{figure}

Finally, we investigate other possibilities for the presence of the gap.  It
is interesting to note in Figure \ref{fig:f12} that at 80~AU, the center of
the gap, the spectral character of the disk is different than that seen in
other portions of the disk.  If this is real, then a different composition in
this annulus may be causing the overall structure of the disk to change, or vice versa.  Gaps in disks tend to receive larger amounts of UV photons due to isotropic scattering of Lyman-$\alpha$ photons from higher layers of the disk \citep{bethel11}.  The excess UV photons can accelerate photo-desorption of ices, which could in turn change the spectral character of the disk within the gap \citep{bergin10,twhyanat}.

Conversely, compositional changes in this gap might be caused by enhanced collisions within 
the disk or the presence of an accreting protoplanet causing a steep truncation to the disk.  Both of these possibilities are supported in part by the recent discovery to a sharp cutoff in the millimeter emission from dust at 60~AU, which is not present in similar high resolution imaging of CO gas \citep{andrews12}.  Since we see scattered light from dust well beyond this cut-off it is clearly due to a change in the bulk size distribution or millimeter emission properties of the dust, which in of itself may be a signature of a planetary companion or change in bulk dust composition.

The gap could instead be an unresolved spiral structure in the disk, as seen in NICMOS images of HD 141569,
where spiral structure at lower spatial resolution mimicked a gap structure
\citep{weinberger99,Clampin}.  This could explain both the asymmetry as well as the
decrease in surface brightness and a hint of non-axisymmetric structure is seen in the higher spatial resolution STIS images of the TW Hya disk.  Higher spatial resolution images might further solve this mystery.

The provocative nature of such a structure indicates that significant, large scale physical processes at large separations occur for protoplanetary disks and can significantly impact their midplane structures.  Whatever the specific origin of this feature within the TW~Hya disk, the combined observations of optical/NIR scattering with longer wavelength photometry and spatially resolved imaging may lead to a new understanding of the conditions in which planetary systems form.

\acknowledgements{The authors would like to thank the anonymous referee for several helpful suggestions, including a suggestion to investigate the spectral type of TW Hya.  We would also like to thank Ted Bergin, Diego Munoz, and Ruobing Dong for enlightening conversations on alternative origins for disk gaps, as well as A. Meredith Hughes for discussions of TW Hya CO data.  Support for program \#10167 was provided by NASA through a grant from the Space Telescope Science Institute, which is operated by the Association of Universities for Research in Astronomy, Inc., under NASA contract NAS 5-26555.}

\bibliography{scibib,twhya}
\bibliographystyle{apj}

\begin{deluxetable}{cccc}
\tablecolumns{4}
\tablewidth{0pc}
\tablecaption{\label{tab:scalings} Photometry of TW~Hya and PSF Scalings}
\tablehead{
\colhead{Filter} & \colhead{PSF Reference Star} &\colhead{Scaling} & \colhead{TW Hya Flux Density (Jy)}
}
\startdata
STIS & HD~85512 & 0.0544,0.0586 & 0.1862 \\
F110W & $\tau^1$ Eridani & 0.099$\pm$0.0027 &  0.747 \\
F160W & Gl~879 & 0.031$\pm$0.002 &  1.03 \\
F171M & CD-43$^\circ$2742 & 0.541$\pm$0.008 &  0.94$\pm$0.01\\ 
F180M & CD-43$^\circ$2742 & 0.528$\pm$0.001 &  0.84$\pm$0.01 \\
F204M & CD-43$^\circ$2742 & 0.56$\pm$0.01 &  0.77$\pm$0.01 \\
F222M & CD-43$^\circ$2742 & 0.61$\pm$0.01 &  0.76$\pm$0.02 \\
\enddata
\end{deluxetable}

\begin{deluxetable}{ccccc}
\tablecolumns{5}
\tablewidth{0pc}
\tablecaption{\label{tab:spectype} Spectral Comparisons from the STIS NGSL}
\tablehead{
\colhead{Star} & \colhead{Spectral Type} &\colhead{T$_{eff}$ (K)} & \colhead{Radius (R$_\odot$)} & \colhead{Refs.}
}
\startdata
HR 8086 & K7V & 3985$\pm$115 & 0.628  & 4,7,8,11,12,16,17 \\
HIP 32939 & M0V & 4352 & ... & 1 \\
GJ 825 & M0V & 3729$\pm$164 & 0.51 &  7,10,14,15 \\
HD 33793 & sdM1 & 3585$\pm$93 & ... & 3,7,14 \\
BD+442051 & M2V & 3600$\pm$300 & ... & 2,6,7,9,11,13,14\\
HD 95735 & M2V & 3597$\pm$131 & 0.359 & 3,5-7,12,16 \\
\enddata
\tablerefs{(1) \citet{ammons06}; (2)\citet{frasca09}; (3) \citet{houdebine10}; (4) \citet{ivanov04}; (5) \citet{jenkins09}; (6) \citet{katz}; (7) \citet{koleva12}; (8) \citet{kovtyukh03}; (9) \citet{milone}; (10) \citet{morales08}; (11) \citet{prugniel01}; (12) \citet{prugniel11}; (13) \citet{soubiran08}; (14) \citet{soubiran10}; (15) \citet{takeda07}; (16) \citet{vanbelle}; (17) \citet{wright11} }
\end{deluxetable}

\begin{deluxetable}{cc}
\tablecolumns{2}
\tablewidth{0pc}
\tablecaption{\label{tab:fitparams} Best-fit Disk Models Values}
\tablehead{
\colhead{Model Parameter} & \colhead{Model Value}
}
\startdata
$a_{\rm min}$ & 0.005~$\mu$m \\
Gap Width &  20~AU \\
Gap Depth & 0.3 \\
Truncation & 100~AU \\
$\chi^2_{nu}$  & 3.29 \\
\hline
\multicolumn{2}{|c|}{Fixed Parameters} \\
\hline
L$_\star$ & 0.208~L$_\odot$ \\
T$_{eff}$ & 3600~K \\
R$_\star$ & 1.08 R$_\odot$ \\
$\dot{M}$ & 10$^{-9}$ M$_\odot$ yr$^{-1}$ \\
$\alpha$ & 5$\times$10$^{-4}$ \\
$a_{\rm max}$ & 1~cm \\
\enddata
\end{deluxetable}

\end{document}